# Hierarchical Serpentine-like Organic Crystal Optical Waveguides for Artificial Neural Networks

Avulu Vinod Kumar,[1] Mehdi Rohullah,[1] Melchi Chosenyah,[1] Sinduja Gaddam,[1] Rajadurai Chandrasekar[1,✉] (ORCID: 0000-0003-3527-3984)

[1]Advanced Photonic Materials and Technology Laboratory, School of Chemistry and Centre for Nanotechnology University of Hyderabad, Prof. C. R. Rao Road, Gachibowli, Hyderabad 500046, Telangana, India.

E-mail: r.chandrasekar@uohyd.ac.in

**ABSTRACT**

Optical components and circuits that deal with multiple signal generation and processing are quintessential for artificial neural networks. Herein, we present a proof-of-concept four-layered organic optical artificial neural network (ANN)-like architecture, constructed from flexible organic crystals of (*E*)-1-(((5-methylpyridin-2-yl)imino)methyl)naphthalene-2-ol (MPyIN), employing an atomic force microscopy cantilever tip-based mechanical micromanipulation technique. Initially, the strategic selection of four MPyIN crystal active waveguides of varying lengths, mechanically bending them into serpentine-like forms, followed by their hierarchical integration, creates neuron-like, four-layered interconnected optical waveguides with six optical synapses. The synapses in the ANN-like architecture enable parallel transmissions of passive optical signals via evanescent coupling across multiple paths through various layers of the serpentine-shaped optical waveguides. Notably, the feedforward mechanism allows the synapses to multiply and split the optical signal generated at any input into four diverging signals with varying magnitudes. Here, certain outputs deliver a mixed signal (passive and active) due to diverging and converging optical transmission paths. This hierarchical, ANN-like tiny architecture paves the way for the development of smart optical neural networks utilizing multiple emissive and phase-changing organic crystals.

**INTRODUCTION**

Understanding the science behind the simultaneous generation and processing of multiple signals is pivotal to unveiling the neural signal processing and cognitive processes in the human brain.[1-2] Inspired by the biological brain, artificial neural networks (ANN) process the input data using several interconnected neurons (nodes).[3-5] Here, each connection acts as a synapse akin to the biological brain. Experimentally, ANN processing can be implemented via photonic integrated circuits.[6,7] For example, systematically arranged optical waveguides (interconnects), producing, transporting, and splitting multiple optical signals become instrumental in devising ANN.[8,9] Silicon-based ANN has its challenges, such as (i) integrating a strong non-linear optical (NLO) material to a weak NLO silicon material, (ii) difficulty in monitoring the feedback operations, network training, node failures, and susceptibility to environmental fluctuations.[10]

These demerits can be avoided using organic materials with superior solid-state opto-mechanical properties.[11,12] Unlike silicon waveguides, which act as passive-only (deliver the same input light at the other end) signal transducers, organic crystals act as active (transduces generated photoluminescence) and passive waveguides depending on the input excitation wavelength.[13-16] Moreover, the long pass filter effect originated from fluorescence (FL) reabsorption and radiative energy transfer (ET) due to the overlap of one crystal's emission with the absorption of another, enabling more than one operational mechanism in organic optical circuits.[15,16] Optical non-linearity, bandwidth tunability, high photoluminescence quantum yield, lightweight, and mechanical flexibility towards integration are added advantages to organic crystals.[17-21] The interconnected organic crystal optical waveguides are useful for ANN. Liao *et al.* demonstrated the utility of epitaxially grown organic crystalline heterostructures for photonic applications.[22,23] Zhang *et al.* fabricated two-dimensionally inter-woven flexible crystal optical waveguides.[24] Our group created a three-dimensionally stacked organic crystal waveguide structure.[25]

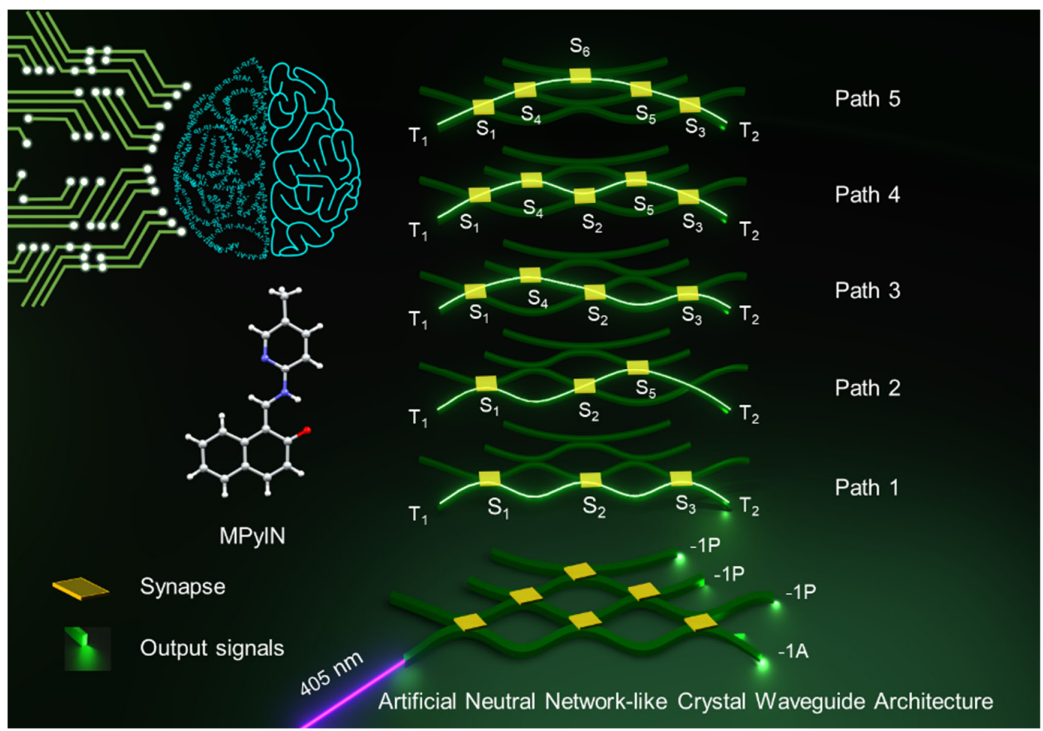

**Fig. 1 | Serpentine-like organic crystal waveguides as artificial neural network (ANN).** Schematic illustration of optical artificial neural network fabricated using hierarchically arranged flexible, serpentine-shaped organic molecular crystal waveguides of MPyIN. The down-left inset shows the molecular structure of MPyIN.

The starting step towards accomplishing ANN from organic crystals is via the fabrication of complex optical trajectories using an optical directional coupler (DC).[26,27] A DC splits the input signal into two or more in different directions. Optical waveguides with curved geometries are necessary to construct DCs. In silicon photonics, waveguide's convex bends are cleverly utilized for the construction of complex DCs, which are imperative for the design of optical ANN technologies.[28] The light traveling at the convex region of the curved waveguide can be effectively coupled to a nearby waveguide via evanescent coupling.[26-29]

Recently, our group reported the *pseudo-plasticity* (substrate-assisted permanent mechanical deformation) in elastic organic crystals self-assembled on a borosilicate substrate.[25] The same property enabled the construction of monolithic and hybrid 2×2 DCs using atomic force microscopy (AFM) cantilever tip-aided *mechanophotonic* approach.[11,28] Despite the successful demonstration of these DCs, the fabrication of organic crystal-based ANN remained an elusive target as it requires many features, including (i) Crystals that can demonstrate pseudo-plasticity when subjected to multiple mechanical bends, (ii) Crystals that can efficiently guide both active and passive optical signals, functioning similar to neurons, (iii) Micromanipulation for the precise positioning of the crystal convex bends to enable evanescent coupling between waveguides, (iv) hierarchical assembly of multiple crystal waveguides in an ordered fashion to create multiple optical synapses (Fig. 1), and (v) Multiple transmissions of optical information from one waveguide to other interconnected waveguides through synapses, forming a complex optical neural network.

Here, we report for the first time an organic crystal-based ANN-like architecture from hierarchically arranged multiply bend waveguides (Fig. 1). We designed a green-FL Schiff-base compound, (*E*)-1-(((5-methylpyridin-2-yl)imino)methyl)naphthalen-2-ol, MPyIN. The weak intermolecular interactions in the molecular crystals provide mechanical flexibility to MPyIN crystals. Firstly, the mechanical micromanipulation of MPyIN crystals allows the construction of a strained waveguide with five curves. Later, the physical integration of convex bends with adjacent MPyIN waveguide generates three optical synapses ($S_1$-$S_3$, evanescent coupling points), which communicate signals from one neuron (waveguide) to another. Furthermore, hierarchically integrating two more bent MPyIN waveguides creates synapses $S_4$-$S_5$ and $S_6$, resulting in a four-level organic crystal-based ANN-like structure. The constructed ANN-like structure simultaneously generates active and passive optical signals with variable split ratios, traveling through four-level hierarchically placed waveguides. Such photonic systems serve as a guide for constructing futuristic organic crystal optical ANNs.

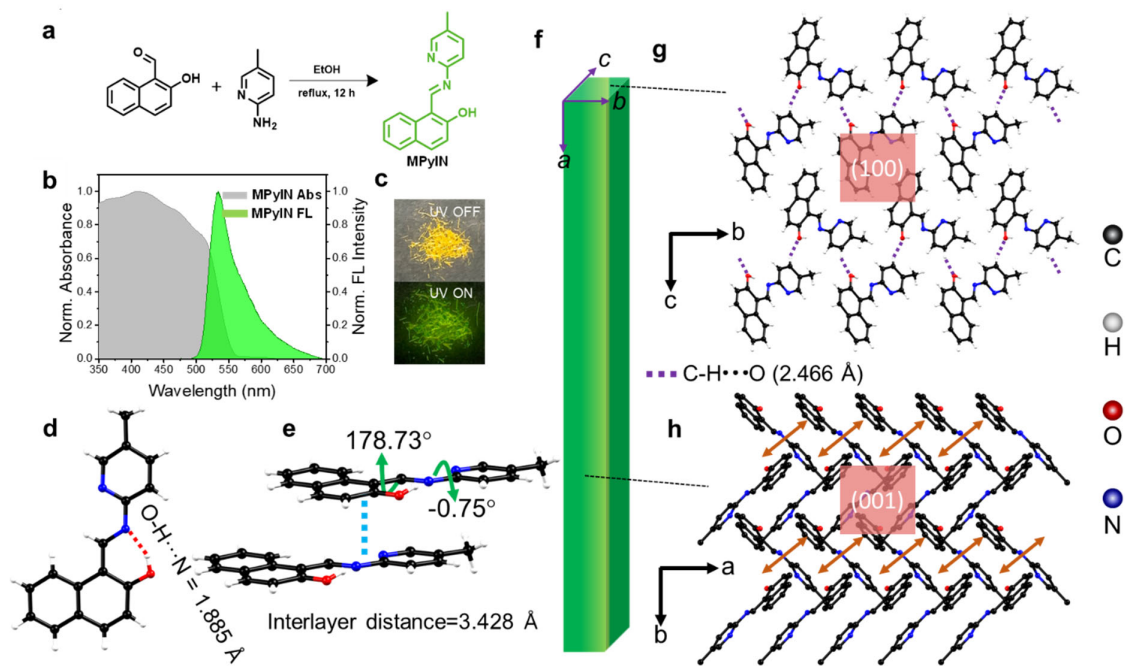

**Fig. 2 | Preparation, photophysical properties, and Single-crystal X-ray diffraction analysis of MPyIN. a** Synthesis of MPyIN molecule. **b** Solid-state photophysical properties of MPyIN molecule. **c** Photographs of needle-like crystals of MPyIN under normal and UV light. **d, e** Top and side views of MPyIN molecule displaying intramolecular hydrogen bonding and torsion angles, respectively. **f** Graphical representation of rectangular morphology of MPyIN crystal. **g, h** Crystallographic *a* and *c* axis showing criss-cross molecular packing of π-π stacked chains.

## RESULTS AND DISCUSSION

The desired MPyIN molecule was synthesized via the well-known Schiff base reaction (Fig. 2a and supplementary Figs. 1-3). The optical absorption of MPyIN extended to ≈ 560 nm from the UV green region (Fig. 2b), whereas its FL ranges from ≈ 490–700 nm with a $\lambda_{max}$ at 540 nm. Here, the absorption band overlaps with its FL band, thereby facilitating reabsorption, which is a crucial phenomenon required for the long-pass filter effect (Fig. 2b).

Green acicular crystals of variable lengths are formed when MPyIN molecules are crystalized in methanol (Fig. 2c). The prerequisite for the fabrication of organic optical components is mechanical flexibility and efficient light guiding in bent geometries. The single crystal X-ray structure analysis of acicular MPyIN crystals with a rectangular morphology (Fig. 2f) revealed the monoclinic crystal system in the $P2_1$ space group (CCDC number: 2414176; Supplementary Table 1). MPyIN molecules in their crystal lattice have intramolecular O-H⋯N hydrogen bonding (1.885 Å, top view) and π⋯π interactions (3.428 Å, side view). As a result, the molecule demonstrated nearly planar geometry with minimal torsion angles between pyridine and naphthalene rings (-0.75° and 178.73°) (Fig. 2d, e). When observed along the crystallographic *c*-axis, the molecules adopted a criss-cross arrangement involving π⋯π interactions (Fig. 2h). To facilitate a better understanding of the crystal's geometry, the primary facets of the mounted macrocrystal were determined to be (001), (010), and (100) through face indexing (Fig. 2g, and Supplementary Fig. 4). The mechanical flexibility of the MPyIN crystal was investigated using the three-point bending technique. In this approach, an external force was applied to the crystal using a thin needle, oriented perpendicular to the thicker (001) plane. The crystal exhibited an elastic response (Supplementary Fig. 5).

Organic ANN-like structure construction entails miniature optical waveguides. Typically, drop-casting a 25 μL of methanolic solution (0.1 mM) of MPyIN on borosilicate coverslip under slow evaporation conditions (used for bulk crystallization) produces microcrystals. The field emission scanning electron micrographs (FESEM) reveal the rectangular morphology of several naturally bent microcrystals (Supplementary Fig. 6). To explore their mechanophotonic behavior, an MPyIN microcrystal (OW1; L ≈ 273 μm; W ≈ 1.65 μm) was chosen (Fig. 3a). The complete illumination of OW1 with 365 nm light revealed a dull body with radiant tips, pointing at the active optical waveguiding capability (Fig. 3a). Therefore, the left terminal $T_1$ of OW1 was irradiated with a 405 nm diode laser, generating a bright green fluorescence ($\lambda_1 \approx$ 475 – 700 nm) at $T_1$. This FL ($\lambda_1 \approx$ 475 – 700 nm) was guided through the crystal and outcoupled at $T_2$ as a reabsorbed signal with a narrow spectral bandwidth ($\lambda_2 \approx$ 540 – 700 nm) (Supplementary Fig. 7 a-c, h). The FL signal was recorded at $T_2$ (Supplementary Fig. 7c), by optically exciting the crystal at multiple positions on OW1. The gradual rise in FL signal intensities, as the optical path length of the propagating light decreases, enabled us to assess the optical loss in the straight OW1. The optical loss of the waveguide was estimated by fitting the plot of $I_{tip}/I_{body}$ versus the propagation length D using the equation, $I_{tip}/I_{body} = e^{-\alpha D}$, where α is the optical loss coefficient, expressed in dB μm$^{-1}$, $I_{body}$ is FL intensity at each excitation position, and $I_{tip}$ is FL intensity at $T_2$ (Supplementary Fig. 7j). From the fit, the value of α', optical loss in dB μm$^{-1}$ (dB loss = 4.34 α) was estimated to be 0.14339 dB μm$^{-1}$ for straight OW1.

The feasibility of organic ANN with MPyIN is possible only if the microcrystals can withstand external stress and perform optical waveguiding in the strained state. Therefore, the straight OW1 was subjected to mechanical force with an AFM cantilever tip under the observation of a confocal microscope at multiple positions by slowly pushing OW1 in ±x and ±y directions. After removing the mechanical force, the crystal retained its bent geometry, which is accredited to the pseudoplastic behavior (Fig. 3b and Supplementary Fig. 7d-f, i). The crystal was intelligently bent in such a way that it induced multiple concave and convex curves on the waveguide. These bends offer effective coupling of optical signals into other waveguides. The bending-induced strain (ε) was calculated using the equation, $\varepsilon = t/2r \times 100$, where t and r are the thickness and radius of the hypothetical circle formed around the bent region of the microcrystal. The confocal image clearly showed that the crystal with five curved bends, C1-C5 upholds a strain (ε) amounting to about 1.7%, 1.7%, 1.3%, 1.7%, and 1.5% at bends b1-b5, respectively (Supplementary Fig. 7g). Experiments performed akin to straight waveguides exhibited the optical waveguiding property of multiply bent OW1 (Supplementary Fig. 7d-f) with an $\alpha' = 0.15910$ dB $\mu m^{-1}$ (Supplementary Fig. 7j). These three convex bends can be strategically utilized to create synapses (S), allowing them to couple light into other optical waveguides, effectively mimicking the behavior of neurons.

To create a four-layered ANN-like structure, four microcrystals must be sequentially integrated at a single location on the substrate. Therefore, three other MPyIN microcrystals, labeled OW2 to OW4, were mechanically moved near the bent OW1 microcrystal and aligned parallel to it using the AFM cantilever tip (Fig. 3a, b). Later, the straight OW2 (L ≈ 306 μm; W ≈ 1.9 μm; $\alpha' = 0.13232$ dB $\mu m^{-1}$) was pushed towards the bent OW1 (Fig. 3b and Supplementary Figs. 8a-c, 9) and diligently bent to create three convex bend facing OW1 (Fig. 3b, c). The challenge here is to seamlessly integrate these two microcrystals so that the evanescent contact region functions as an optical synapse, similar to the neural network. Through meticulous mechanical maneuver, their convex sections were brought into physical contact to form a 2×2 DC, enabling effective evanescent coupling between the OW1 and OW2 waveguides at synapses, $S_1$-$S_3$ (Fig. 3c and Supplementary Fig. 8c, d and Supplementary video 1). Thereby, OW2 serves as the second layer for the organic ANN structure.

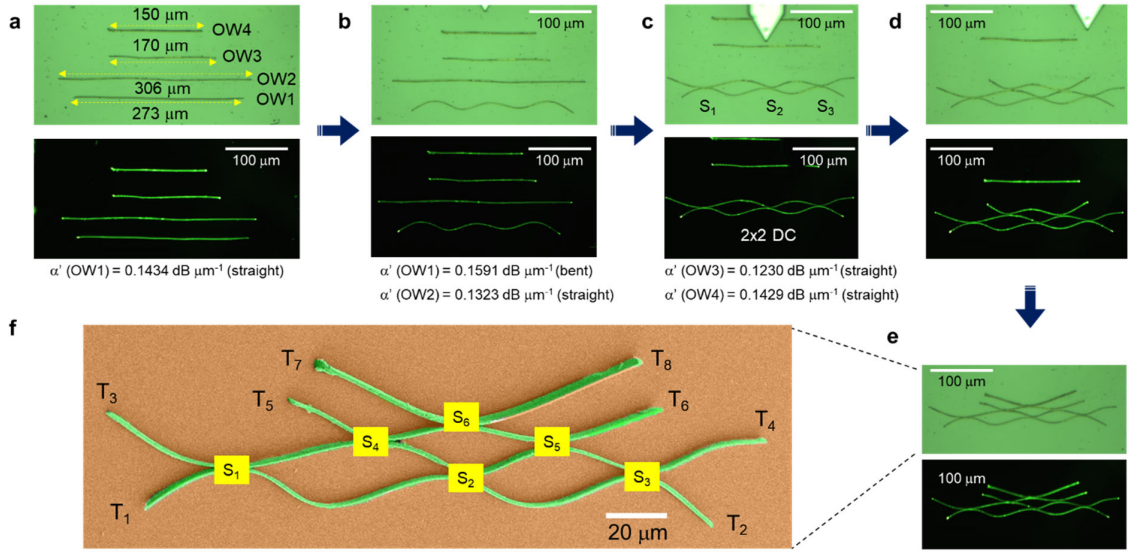

**Fig. 3 | Fabrication of four-layered ANN using MPyIN microcrystals.** Confocal optical and FL images of **a** four straight single crystal optical waveguides OW1 – OW4 arranged parallel to each other, **b** mechanically bent OW1, **c** bending of straight OW2 and creation of 2×2 directional coupler (DC), by integrating bent OW1 and OW2. **d** Micromechanical bending and integration of OW3 with OW2 to form a three-layered monolithic ANN. **e** Bending of straight OW4 and its integration with OW3 to fabricate a four-layered ANN-like architecture. **f** Color-coded FESEM image displaying the fabricated four-layered ANN-like architecture with six synapses.

The optical signal transmission abilities of the constructed two-layered 2×2 DC with three synapses were investigated by introducing input light at one terminal and observing the optical signals at the other terminals (Fig. 4a). For instance, terminal $T_1$ (in the first layer of ANN) was stimulated with a 405 nm laser, resulting in the observation of bright green FL (1A), which actively propagated to output terminal $T_2$ in bent OW1 (Fig. 4b). As 1A traveled towards terminal $T_2$, a portion of the shorter wavelength region was filtered out due to reabsorption, and delivered signal -1A (where minus represents reabsorbed signal and A denotes active) at $T_2$ (Fig. 4d, f). While delivering the optical signal to terminal $T_2$, some of the light was

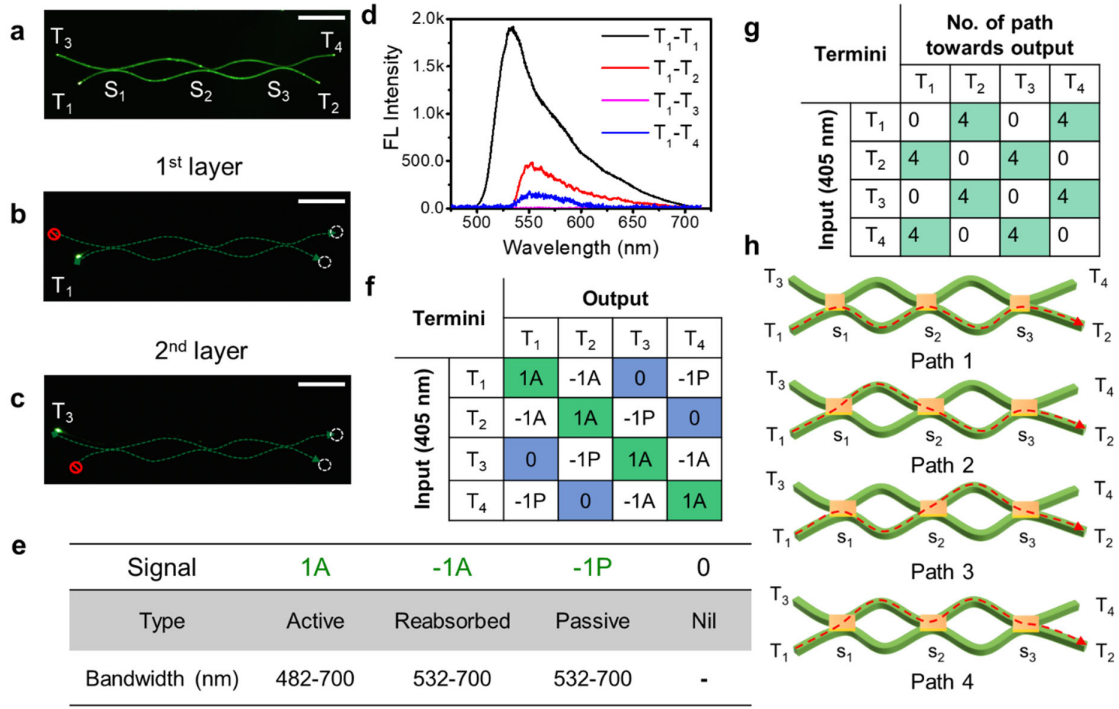

**Fig. 4 | Optical performances of 2×2 Directional Coupler (DC). a** FL image of 2×2 directional coupler (DC), with three synapses from multiply bent OW1 and OW2. **b,c** FL images of DC excited at $T_1$ and $T_3$ terminals. The scale bar is 50 μm. **d** FL spectra recorded at other termini when excited at $T_1$. The green dotted lines in FL images indicate the direction of light propagation in the 2×2 DC. White circles point at the outcoupled light in the waveguides. **e** Table depicting different types of optical signals with their respective bandwidths. **f** Optical signals observed at various output termini are tabulated against the given input. **g** Table depicting the number of paths taken for optical signals to reach various output termini are tabulated against the given input. **h** Graphical representation of 2×2 DC with three synapses showing four different paths (red-dotted arrows) towards $T_2$ for input at $T_1$.

channeled into OW2 through optical synapses ($S_1$-$S_3$) via evanescent coupling and passively propagated towards $T_4$ as -1P, where P signifies passive signal (Fig. 4d, f, h). Unlike the previously known DCs,[25,26] the notable aspect of this DC is that it has multiple optical coupling regions, $S_1$-$S_3$. As a result, the split passive light reenters OW1 two times through $S_2$ and $S_3$, ensuring efficient mixing of guided active and passive lights. The DC configuration also forbids signals reaching $T_3$, indicated as 0 (Fig. 4b, d, f). Due to three synaptic connections, four different optical pathways are available for the signal to reach $T_2$ for input at $T_1$ (Fig. 4g, h), analogous to the behavior of neurons. Path 1: $S_1$(-1A)→$S_2$(-1A)→$S_3$(-1A); Path 2: $S_1$(-1P)→$S_2$(-1P)→$S_3$(-1P); Path 3: $S_1$(-1A)→$S_2$(-1P)→$S_3$(-1P); Path 4: $S_1$(-1P)→$S_2$(-1P)→$S_3$(-1P). Similarly, for each input at terminals $T_2$, $T_3$, and $T_4$, the corresponding outcoupled optical signals at the other terminals followed four distinct optical transmission paths (Fig. 4f-h, Supplementary Figs. 10, 11)

To enhance the functional complexity of the organic ANN with a three-layered structure and five synapses, an additional optical waveguide (OW3; L ≈ 170 μm; W ≈ 3.3 μm; α' = 0.12303 dB μm$^{-1}$) was doubly bent and integrated with two convex regions of OW2 using micromechanical manipulation strategies (Fig. 5a, and Supplementary Figs. 8e, f, and 12). During this process, two additional synapses, $S_4$ and $S_5$, were formed, enabling extended light flow within the fabricated optical network consisting of OW1 to OW3 with five synapses (Fig. 5a-d). Here, the introduction of light at any input terminal effectively produced three outputs with varying signal split ratios transducing through distinct optical paths (Fig. 5e,f). When input was given at $T_1$, the active signal (1A) transduced towards $T_2$ and outcoupled as -1A (Fig. 5b, e and Supplementary Fig. 13). Further, the propagating 1A signal coupled into adjacent OW2 at $S_1$, $S_2$, and $S_3$ to yield -1P signal at $T_4$ (Fig. 5b, e and Supplementary Fig. 13). Further, some of the light from OW2 passed onto the remotely placed OW3 through $S_4$ and $S_5$, resulting in a -1P signal at $T_6$ (Supplementary Fig. 13). However, the optical signal intensity at $T_6$ was weak ascribed to two indirect transfers from OW1. Similarly, light processing at other terminals is tabulated in Fig. 5b-e. Due to the presence of five synapses ($S_1$-$S_5$), the number of diverging and converging optical pathways varied from two to five depending on the input point. For example, with input at $T_1$, five optical trajectories route the signal to $T_2$ as (Fig. 5f, g and Supplementary Fig. 13): Path 1: $S_1(-1A) \rightarrow S_2(-1A) \rightarrow S_3(-1A)$; Path 2: $S_1(-1A) \rightarrow S_2(-1P) \rightarrow S_5(-1P) \rightarrow S_3(-1P)$; Path 3: $S_1(-1P) \rightarrow S_4(-1P) \rightarrow S_2(-1P) \rightarrow S_3(-1P)$; Path 4: $S_1(-1P) \rightarrow S_4(-1P) \rightarrow S_2(-1P) \rightarrow S_5(-1P) \rightarrow S_3(-1P)$; Path 5: $S_1(-1P) \rightarrow S_4(-1P) \rightarrow S_5(-1P) \rightarrow S_3(-1P)$. Similarly, a mixed signal reaches $T_4$, traveling through five optical pathways, while only three paths are required for the signal to reach $T_6$. Likewise, inputs at $T_2$, $T_3$, and $T_4$, transduced signal to outputs through five and three synapses. In contrast, inputs at $T_5$ route the signal through three synapses to $T_2$ and $T_4$, and two synapses to $T_6$. The same applies to $T_6$, though the routing is directed to $T_1$, $T_3$, and $T_5$, respectively (Fig. 5f and Supplementary Figs. 13-18)

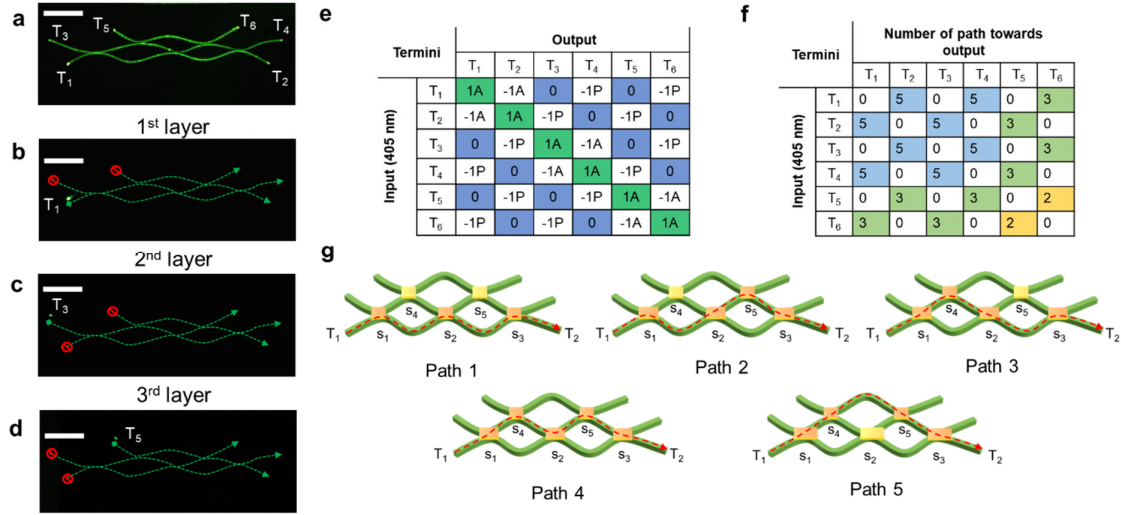

**Fig. 5 | Optical performances of three-layered ANN. a** FL image of three-layered ANN made from multiply bent OW1, OW2, and OW3. **b-d** FL images of ANN excited at $T_1$, $T_3$, and $T_5$ terminals, respectively. The scale bar is 50 μm. **e** Optical signals recorded at various termini against the given input. **f** Table depicting the number of paths taken for optical signals to reach various output termini against the given input. **g** Graphical representation of three-layered ANN with five synapses showing five different paths (red-dotted arrows) towards $T_2$ for input at $T_1$.

To further expand the ANN architecture to four-layers, an additional optical waveguide, OW4 (L ≈ 150 μm; W ≈ 3.3 μm; α' = 0.14291 dB μm$^{-1}$), was singly bent and integrated into the convex section of OW3 forming an optical synapse, S$_6$ (Figs. 3d, e, 5a and Supplementary Figs. 8g, h, 19). Despite the bending-induced strains at each layer, the resulting serpentine-like architecture remained stable after gold coating and under FESEM measurement conditions, except for some minimal disruption to two synaptic contacts (Fig. 3f and Supplementary Fig. 19d). Finally, to understand the optical signal processing capabilities in organic four-layered ANN-like architecture, light was introduced at one terminal, and the optical communication across different layers occurring via six optical synapses at other terminals was monitored. During the propagation of the input optical signal (1A) generated at T$_1$ of OW1 towards T$_2$ as signal -1A, the six synapses split the propagating light into -1P. Additionally, due to optical communication via synapses S$_1$-S$_6$, the passive signal (-1P) was recorded at T$_4$, T$_6$, and at the terminal T$_8$ of the fourth output layer (Fig. 6b, e). However, the output signal intensity gradually reduced from T$_4$ to T$_8$ due to the increasing number of signal transfers through synapses (Supplementary Fig. 19). Light processing for input at various terminals is tabulated in Fig. 6e. Further, careful analysis of the signal pathways of the organic ANN-like structure revealed complex, diverging, and converging, curved optical trajectories. For example, an input generated at T$_1$ revealed five optical trajectories towards outputs T$_2$ through synapses, which are as follows (Fig. 6f, g): Path 1: S$_1$(-1A)→S$_2$(-1A)→S$_3$(-1A); Path 2: S$_1$(-1A)→S$_2$(-1P)→S$_5$(-1P)→S$_3$(-1P); Path 3: S$_1$(-1P)→S$_4$(-1P)→S$_2$(-1P)→S$_3$(-1P); Path 4: S$_1$(-1P)→S$_4$(-1P)→S$_2$(-1P)→S$_5$(-1P)→S$_3$(-1P); Path 5: S$_1$(-1P)→S$_4$(-1P)→S$_6$(-1P)→S$_5$(-1P)→S$_3$(-1P). Moreover, due to these multiple diverging and converging signal propagation pathways, the delivered signal at T$_2$ is a spectroscopically indistinguishable, mixed active and passive signal.

Likewise, for input at T$_3$ and T$_5$, the number of optical paths towards T$_4$, T$_6$, and T$_8$ varied as 5, 3, 1, and 3, 2, 1, respectively (Fig. 6f, and Supplementary Figs. 21, 23) delivering mixed active and passive signals as outputs at $T_4$, $T_6$, and $T_8$. However, the ANN geometry allows the delivery of a guided active signal at $T_8$ for an input given at $T_7$ and pure passive signals at $T_6$, $T_4$, and $T_2$ (Supplementary Fig. 25). Light propagation in different pathways for input at various terminals is tabulated in Fig. 6g (Supplementary Figs. 19-26).

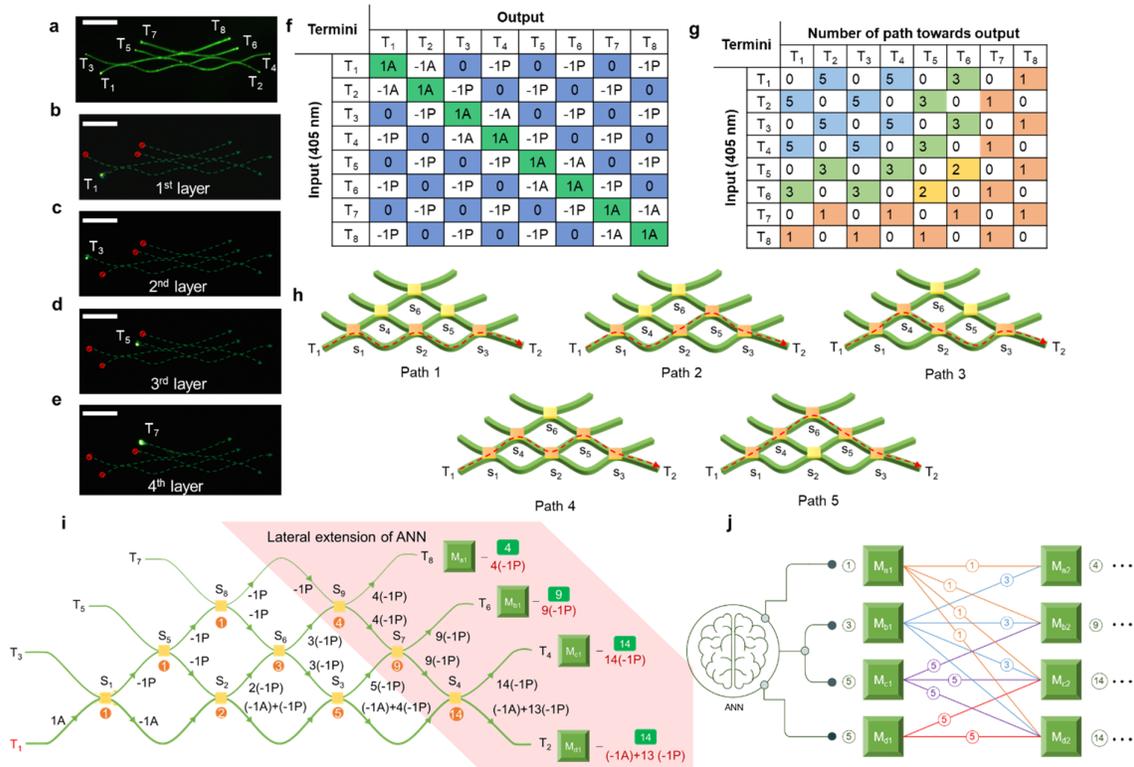

**Fig. 6 | Optical performances and lateral extension of four-layered ANN. a** FL microscopy image of flexible crystal-based four-layered optical ANN-like architecture. **b-e** FL images when laser input was provided at $T_1$, $T_3$, $T_5$ and $T_7$ termini, respectively. **f** Optical performance of organic ANN-like structure. **g** Table depicting the number of paths taken for optical signals to reach various output termini against the given input. **h** Graphical representation of four-layered ANN with six synapses showing five different paths (red-dotted arrows) towards $T_2$ for input at $T_1$. **i** Mixing of signals in laterally extended four-level ANN with an additional three synapses. **j** Satistical prediction of the mixed signals at outputs of the laterally extended ANN.

One of the remarkable features of the multiple optical paths in this constructed four-level ANN is their capacity to split and mix signals at each synapse for any given input. This is possible as each waveguide, which acts as a neuron, can transduce multiple signals. As one can see for the input at $T_1$, outputs $T_2$, $T_4$, $T_6$ and $T_8$ deliver 5 (active and passive), 5 (passive), 3 (passive), and 1 (passive) mixed signals, respectively. Lateral extension of the four-level ANN with an additional three synapses will allow further mixing of the signals in the total of nine synapses and can deliver 14 (active and passive), 14 (passive), 9 (passive), and 4 (passive) mixed signals at outputs $T_2$, $T_4$, $T_6$ and $T_8$, respectively. Similarly, one can statistically predict the complex pattern of mixed signals that will be routed to different outputs for incorporating more lateral synapses (Figure 6i, Supplementary Fig. 27). The ANN structure can be reproduced with the desired number of synapses and waveguides. (Supplementary Fig. 28-31).

This study demonstrated the first fabrication of a proof-of-concept four-layered optical ANN-like architecture using flexible, FL organic crystals of MPyIN by applying the AFM cantilever tip-based mechanical micromanipulation (mechanophotonics) technique. This hierarchical four-layered ANN required four pseudoplastic crystal optical waveguides of different lengths and serpentine-like bends (from three to one). The careful physical integration of the convex bends of each crystal waveguide provided six optical synapses. The structure of organic ANN design closely mimics the edifice of biological neural networks. These synapses functioned as FL light channels, enabling guided transduction of optical signals via evanescent coupling across multiple layers of hierarchically arranged curved crystal waveguides. The key feature of the organic crystal ANN is its operation as a feedforward network. An input at any terminal can generate an active FL signal, split it into multiple propagating active and passive signals through diverging and converging optical paths, and ultimately deliver two optical (active and passive) signals to four out of seven output terminals through a feedforward mechanism. Interestingly, the constructed organic ANN-like architecture enables input-dependent, multiple converging, and diverging guided optical paths (ranging from five to one) through synapses before the mixed (active/passive) or clean (active) signal reaches a specific output. This innovative organic crystal-based optical ANN will serve as a foundation for developing smart and intelligent optical neural networks in the near future, utilizing multiple emissive organic crystals.

**DATA AVAILABILITY**

The X-ray crystallographic coordinates for structures reported in this study have been deposited at the Cambridge Crystallographic Data Centre (CCDC), under deposition numbers 2414176. These data can be obtained free of charge from The Cambridge Crystallographic Data Centre via www.ccdc.cam.ac.uk/data_request/cif.

The data that support the plots within this paper and other finding of this study are available within this article and its supplementary information file, and are also available from the corresponding author upon request.

## ACKNOWLEDGMENTS

RC thanks the DST (DST/INT/RUS/RSF/P-71/2023(G)), SERB (STR/2022/00011 and CRG/2023/003911), and IISc-STARS (MoE-STARS/STARS-2/2023-0465) for funding. MR and MC thanks CSIR for the fellowship.

## AUTHOR CONTRIBUTIONS

RC designed the ANN project. AVK, MR, and MC carried out work under RC's supervision. AVK, MR, and MC equally contributed to the work. SG crystallized the compound under the supervision of RC. All authors discussed the results and wrote the paper.

## COMPETING INTERESTS

The authors declare no competing interest.

**KEYWORDS:** optical Neural Networks • flexible crystal • mechanophotonics • organic directional couplers • optical waveguides • pseudo-plastic crystals

# Supplementary information

# Hierarchical Serpentine-like Organic Crystal Optical Waveguides for Artificial Neural Networks

Avulu Vinod Kumar,[†] Mehdi Rohullah,[†] Melchi Chosenyah,[†] Sinduja Gaddam, and Rajadurai Chandrasekar[*]

Avulu Vinod Kumar,[†] Mehdi Rohullah,[†] Melchi Chosenyah,[†] and Rajadurai Chandrasekar[*]
Advanced Photonic Materials and Technology Laboratory, School of Chemistry and Centre for Nanotechnology
University of Hyderabad, Prof. C. R. Rao Road, Gachibowli, Hyderabad 500 046, Telangana, India
E-mail: r.chandrasekar@uohyd.ac.in

[†] Equal contribution of authors

## Table of Contents:

## 1. Materials

All chemicals and reagents were purchased from commercial suppliers. HPLC solvents were used for synthesis and self-assembly.

## 2. Instrumental Methods

**a) NMR spectroscopy studies:** $^1$H and $^{13}$C NMR spectra were recorded on a Bruker DPX 400 MHz spectrometer with a solvent proton as internal standard (CDCl$_3$: $^1$H: 7.26 ppm, $^{13}$C: 77.16 ppm). Commercially available deuterated CDCl$_3$ was used. Chemical shifts (δ) are given in parts per million (ppm). Spectra were processed using topspin 4.1.1. software.

**b) Absorbance and emission studies:** The measurements were done on a Jasco V-750 spectrophotometer in a diffuse reflectance UV−visible (DR−UV−vis) mode. The reflectance spectra were converted to an absorbance using the Kubelka−Munk function. The solid and solution-state emission spectra were collected using an FP-8500 fluorescence spectrometer. The parameters used where Ex and Em bandwidth are 2.5 nm, Response = 1sec, Sensitivity = Medium, Data interval = 0.5 nm, Ex wavelength = 440.0 nm, Scan speed 500 nm/min.

**c) Single crystal X-ray diffraction:** Single-crystal X-ray diffraction data was collected on Rigaku Oxford XtaLAB ProPilatus3 R 200K-A detector system equipped with a CuKα (λ = 1.54184 Å) MicroMax-003 microfocus sealed tube operated at 50 kV and 0.6 mA. All data were collected at 293 K, and the data reduction was performed using CrysAlisPro software. The crystal structure was refined and solved by using the OLEX software.

**d) Confocal Microspectroscopy studies:** The optical experiments of a single microcrystal were carried out on a transmission mode setup of the Wi-Tec alpha 300 AR laser confocal optical microscope (LCOM) equipped with a Peltier-cooled CCD detector. Using 300 grooves/mm grating BLZ = 750 nm, the accumulation time was adjusted to 10 s and the integration time was typically made 0.5 s. Ten averaged accumulations obtained every single spectrum. A diode 405 nm laser source was used for excitation. 20x objective was used for spectra and image collection, respectively. All the experiments were carried out under ambient conditions.

**e) Micromanipulation of the crystals:** The micromanipulation experiments were performed using the AFM facility attached to the above-mentioned confocal microscope setup. Microcrystals were dispersed on a clean coverslip and dried under ambient condition. Single microcrystals of MPyIN were studied separately. Then each MPyIN microcrystals were transferred onto the circuit fabrication area from different places of the coverslip (borosilicate; Borosil), using an AFM cantilever. An AFM cantilever (TipsNano: NSG10, force constant 3.1−37.6 N/m) was used for the mechanical manipulation. Initially, the AFM cantilever was attached to the holder and aligned to centre by using x/y/z directions control under a confocal microscope. Then the cantilever was moved in –/+z direction to reach the microcrystal. For a typical bending experiment, the AFM-tip was held constant at a height above the piezo stage containing microcrystals, and the piezo stage was carefully maneuvered in +/- x and +/- y directions to bring mechanical deformation in the microcrystal.

**f) Field-Emission Scanning Electron Microscopy:** The morphological analysis was performed using a Zeiss field-emission scanning electron microscope (FESEM) operating at 3 kV. All the experiments were performed after gold coating the samples before imaging.

## 3. Synthesis of MPyIN

The MPyIN was synthesized by the Schiff base reaction between 2-hydroxy-1-naphthaldehyde and 4-methyl-2-aminopyridine in ethanol at 70 ºC to obtain the enol of MPyIN (Scheme S1). 2-hydroxy-1-naphthaldehyde (5 mmol, 1 eq) and 4-methyl-2-aminopyridine (5 mmol, 1 eq) were taken in ethanol in a 100 mL RB flask. The reaction was heated at 70 ºC for 12 h. After cooling to ambient temperature, the precipitate was purified by recrystallization in methanol solvent, yielding green MPyIN needles with a 70% yield.

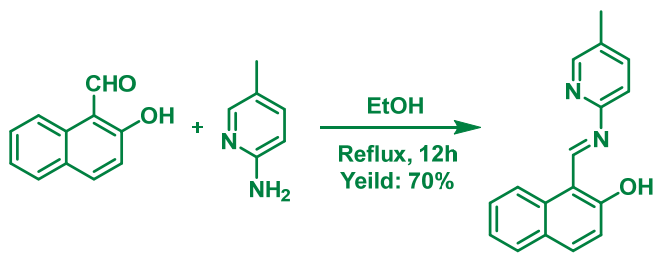

**Supplementary Scheme 1.** Synthesis of MPyIN.

## 4. Preparation of MPyIN microcrystals

Microcrystals were obtained via self-assembly method. In a clean vial, MPyIN (2 mg) was dissolved in methanol (2 mL) and left undisturbed at rt for 6 h. This solution under slow evaporation conditions formed needle-like crystals. Smooth-surfaced microcrystals were formed by drop casting two to three drops (50–80 µl) of MPyIN methanol solution onto a cleaned glass cover slip.

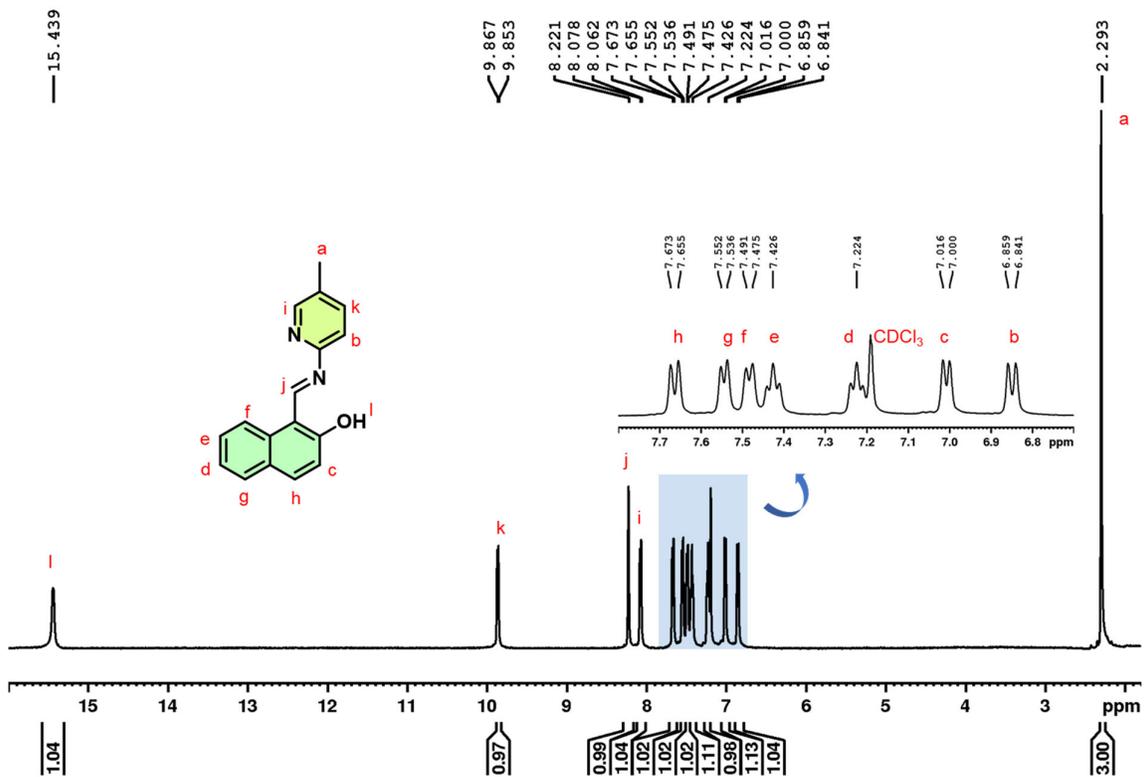

**Supplementary Figure 1.** $^1$H NMR (CDCl$_3$: 400 MHz) spectrum of MPyIN.

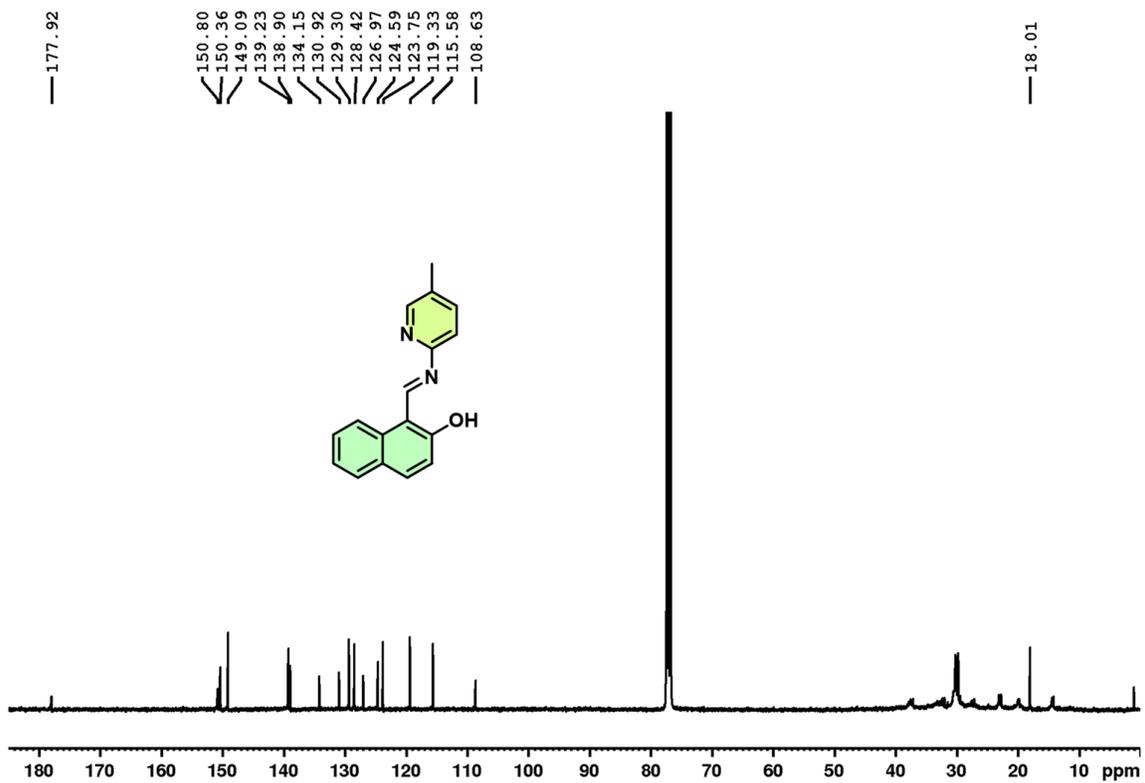

**Supplementary Figure 2.** $^{13}$C NMR (CDCl$_3$: 100 MHz) spectrum of MPyIN.

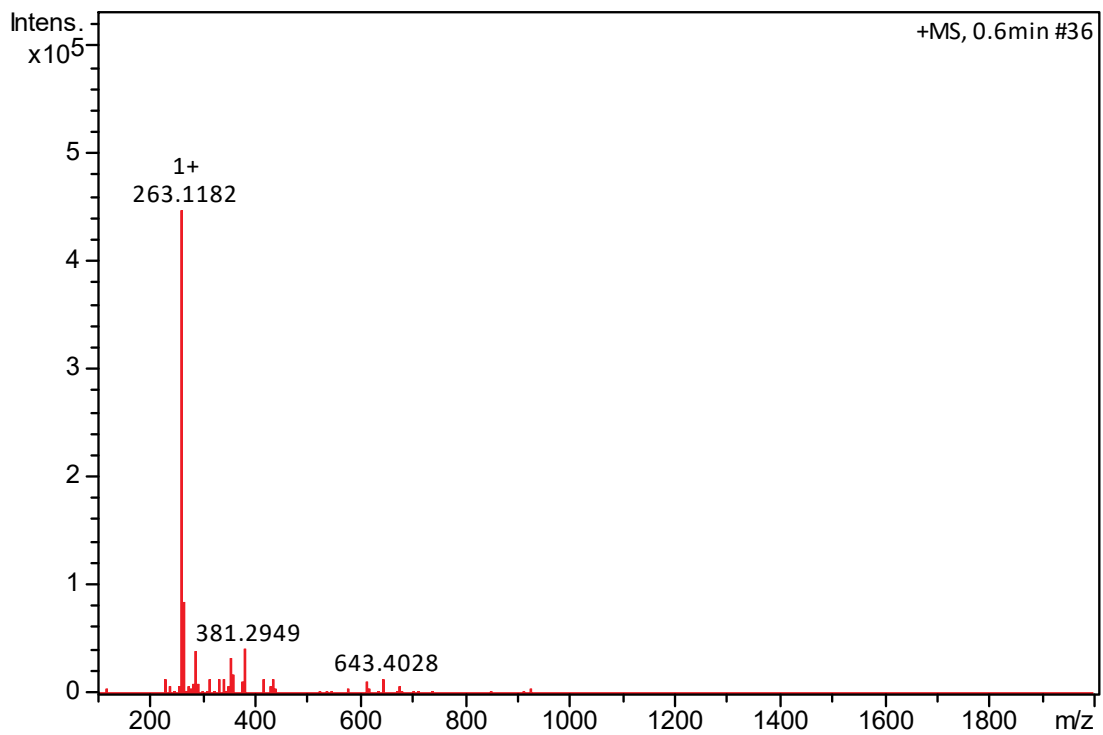

**Supplementary Figure 3.** HRMS (ESI-TOF) spectrum of MPyIN.

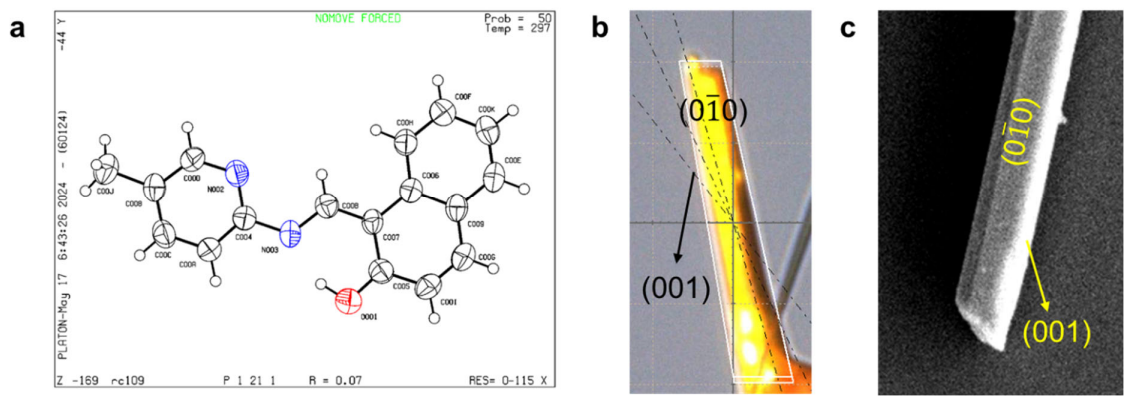

**Supplementary Figure 4. a** ORTEP diagram of MPyIN crystal showing the atoms as thermal ellipsoids plotted at 50% probability. **b** Picture of the mounted MPyIN single crystal. **c** FESEM image of the indexed planes of MPyIN single crystal.

**Supplementary Table 1.** Crystallographic data of MPyIN single crystal.

| Compound name | MPyIN |
|---|---|
| CCDC number | 2414176 |
| Empirical formula | $C_{17}H_{14}N_2O$ |
| Formula weight | 262.31 |
| Temperature (K) | 298 |
| Wavelength | 0.71073 Å |
| Crystal system | Monoclinic |
| Space group | *P*2$_1$ |
| Crystal color | Yellow |
| Cell Lengths (Å) | **a** = 4.8691(4), **b** = 9.5543(10), **c** = 14.0595(17) |
| Cell Angle (°) | **α** = 90, **β** = 98.295, **γ** = 90 |
| Cell Volume (Å³) | 647.217 |
| Z | 9 |
| R-factor (%) | 7.03 |

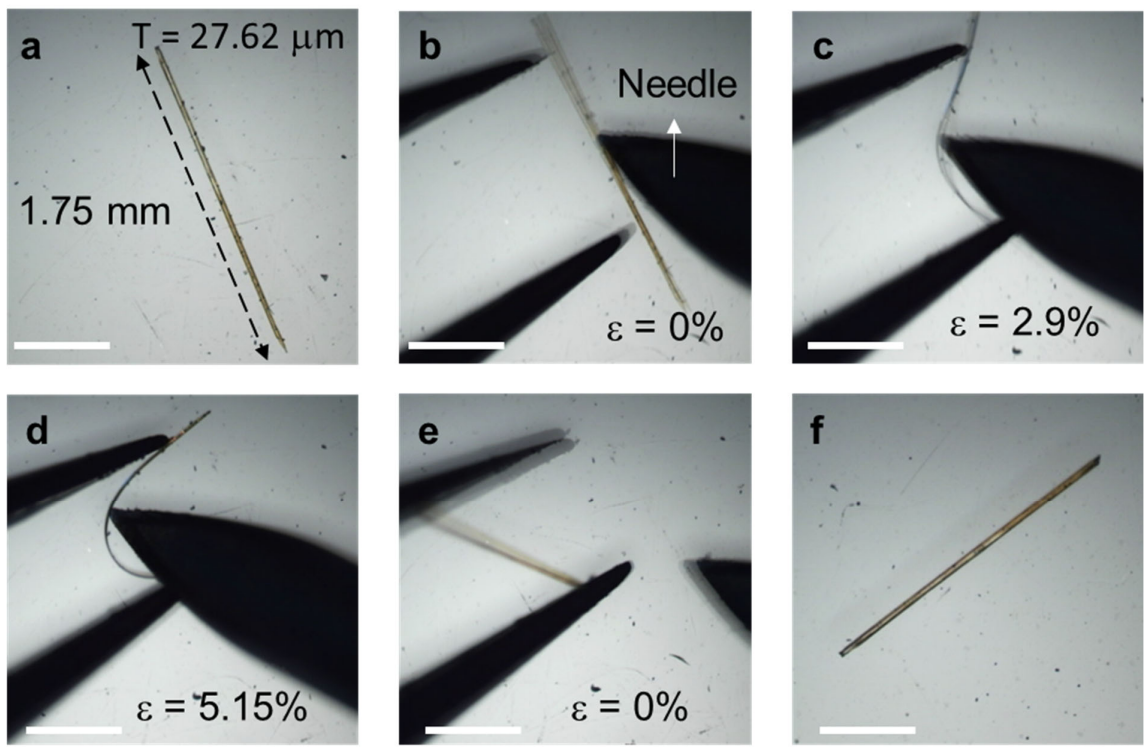

**Supplementary Figure 5.** Three-point bending of MPyIN crystal. The scale bar is 1 mm.

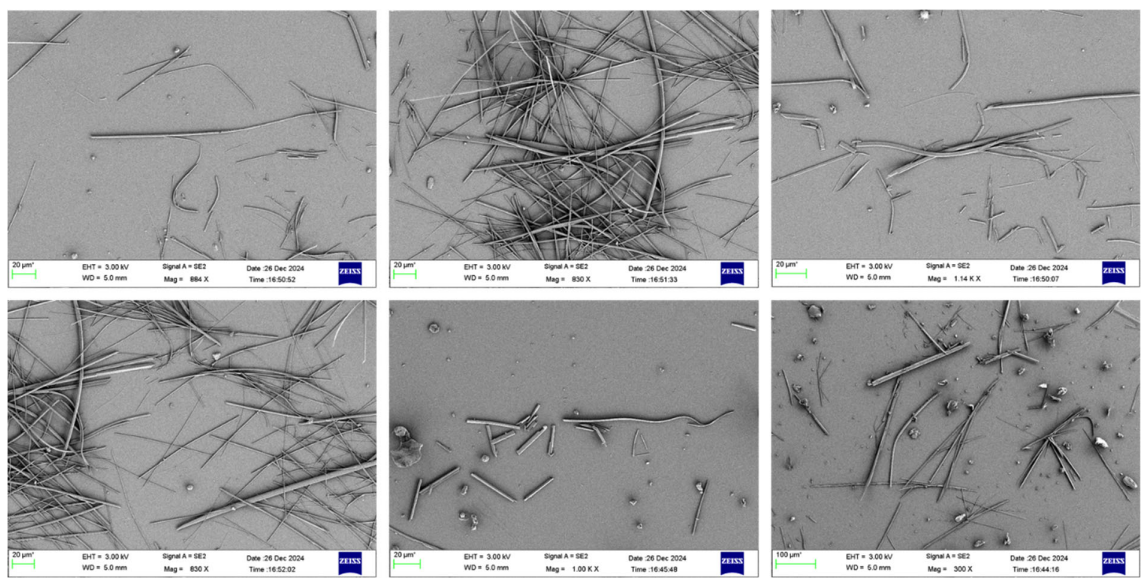

**Supplementary Figure 6.** FESEM images of MPyIN microcrystals. The scale bar is 20 μm.

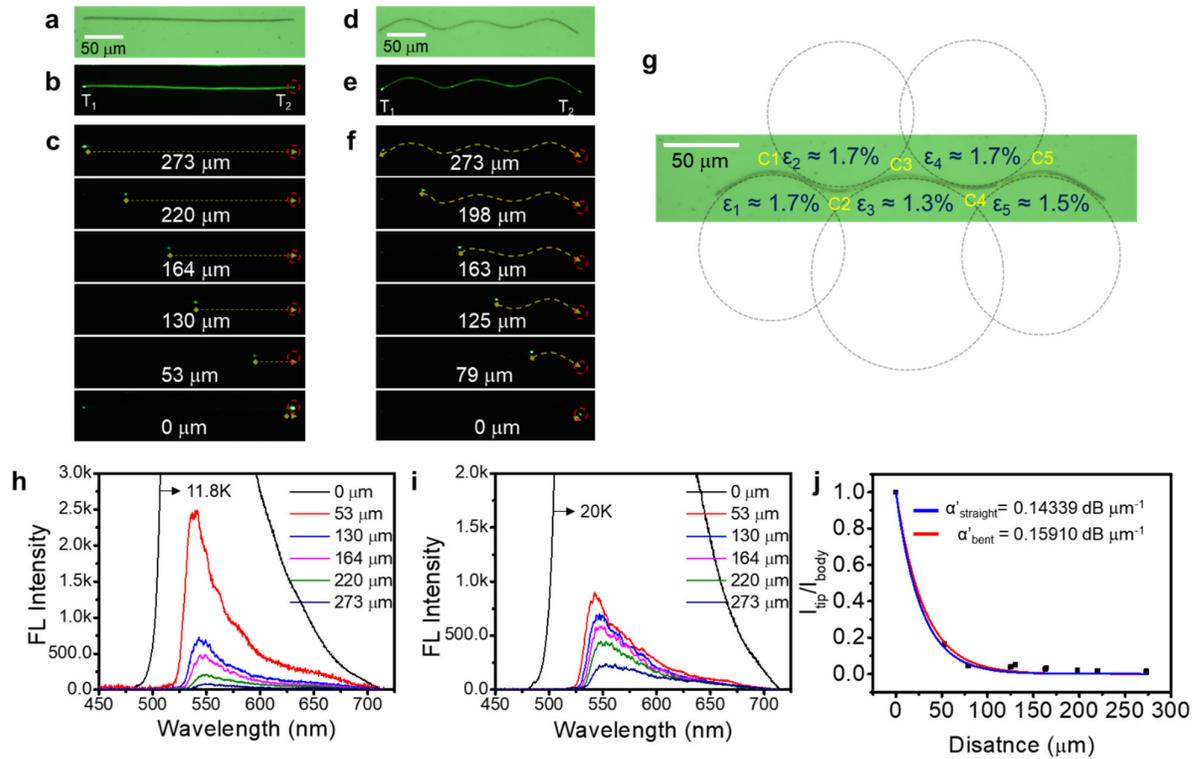

**Supplementary Figure 7.** Confocal optical and FL microscope images of **a, b, c** straight and **d, e, f** mechanically bent MPyIN crystal waveguide (OW1). **g** strain calculation at the curved regions in bent OW1 micromanipulated using AFM cantilever tip. FL spectra were collected at the right end $T_2$ for multiple optical path lengths in **h** straight and **i** bent OW1. **j** corresponding optical loss plots. The $\alpha'$ represents the optical loss coefficient for respective waveguides.

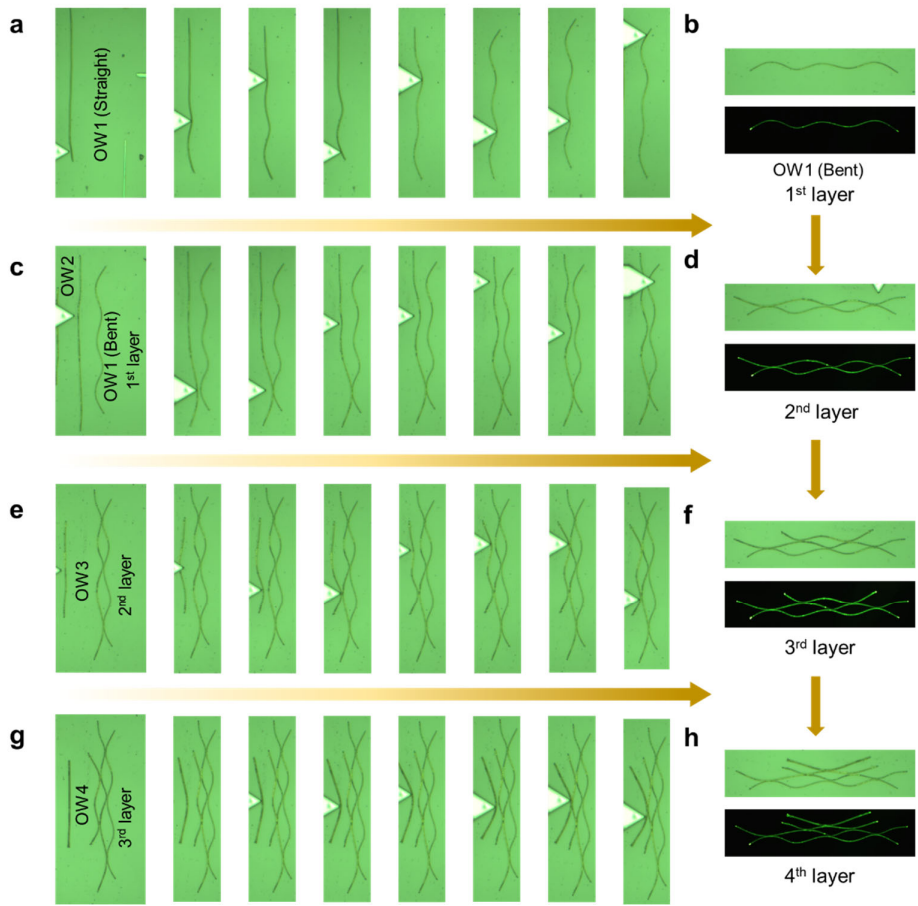

**Supplementary Figure 8. a-h** Confocal optical images of mechanical micromanipulation using AFM tip to fabricate four-layered ANN.

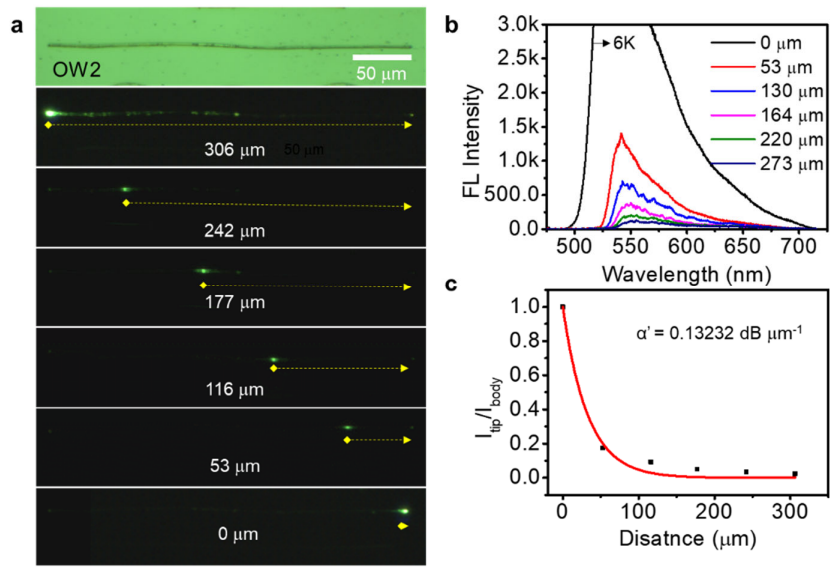

**Supplementary Figure 9. a** Confocal optical and FL microscope images of straight OW2. **b** FL spectra collected at the right end of the waveguide for different optical path lengths in OW2. **c** corresponding optical loss plot.

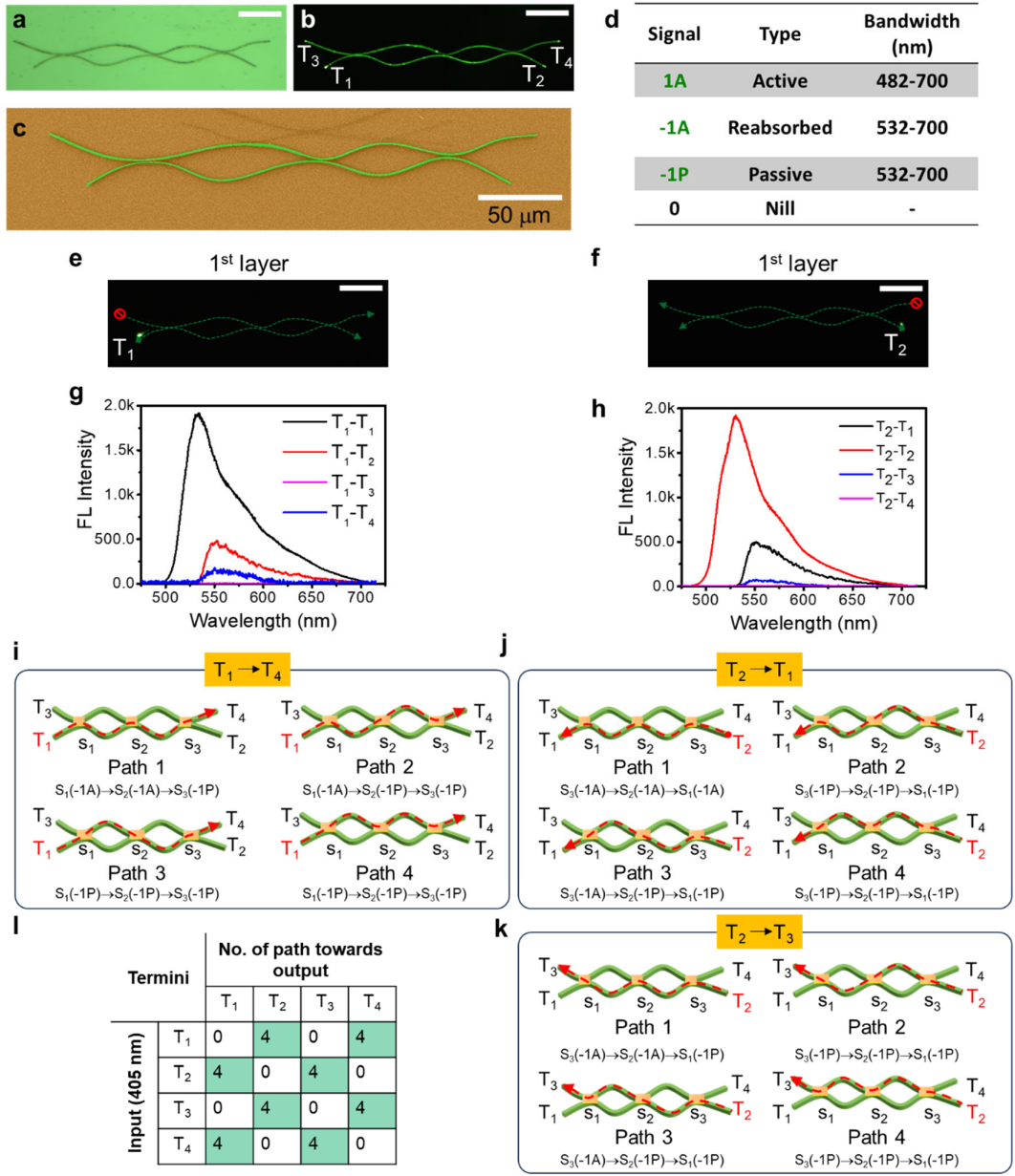

**Supplementary Figure 10. a, b** Confocal optical and FL microscope images of the two-layered ANN. The scale bar is 50 μm. **c** False color coded FESEM image of two-layered ANN. **d** Table depicting different types of optical signals with their respective bandwidths. **e, f** FL images of two-layered ANN excited at $T_1$ and $T_2$ terminals. The scale bar is 50 μm. **g, h** Corresponding FL spectra collected at all the terminals. Graphical representation of two-layered ANN with three synapses showing different paths (red-dotted arrows) **i** towards $T_4$ for input at $T_1$ **j** towards $T_1$ for input at $T_2$ **k** towards $T_3$ for input at $T_2$. **l** Table depicting the number of paths taken for optical signals to reach various output termini against the given input.

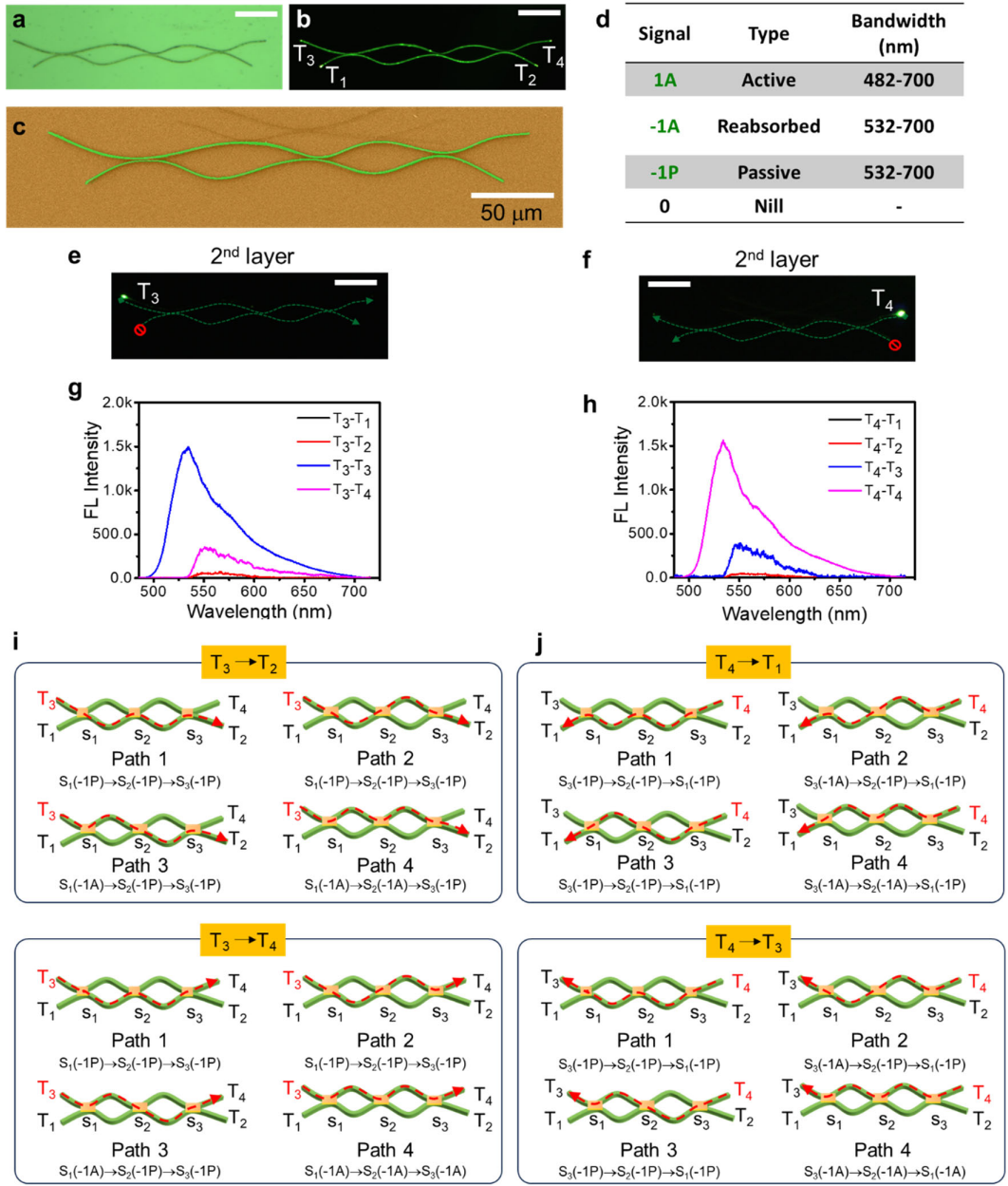

**Supplementary Figure 11. a, b** Confocal optical and FL microscope images of the two-layered ANN. The scale bar is 50 μm. **c** False color coded FESEM image of two-layered ANN. **d** Table depicting different types of optical signals with their respective bandwidths. **e, f** FL images of two-layered ANN excited at $T_3$ and $T_4$ terminals. The scale bar is 50 μm. **g, h** Corresponding FL spectra collected at all the terminals. Graphical representation of 2×2 DC with three synapses showing different paths (red-dotted arrows) **i** towards $T_2$, and $T_4$ for input at $T_3$ **j** towards $T_1$, and $T_3$ for input at $T_4$.

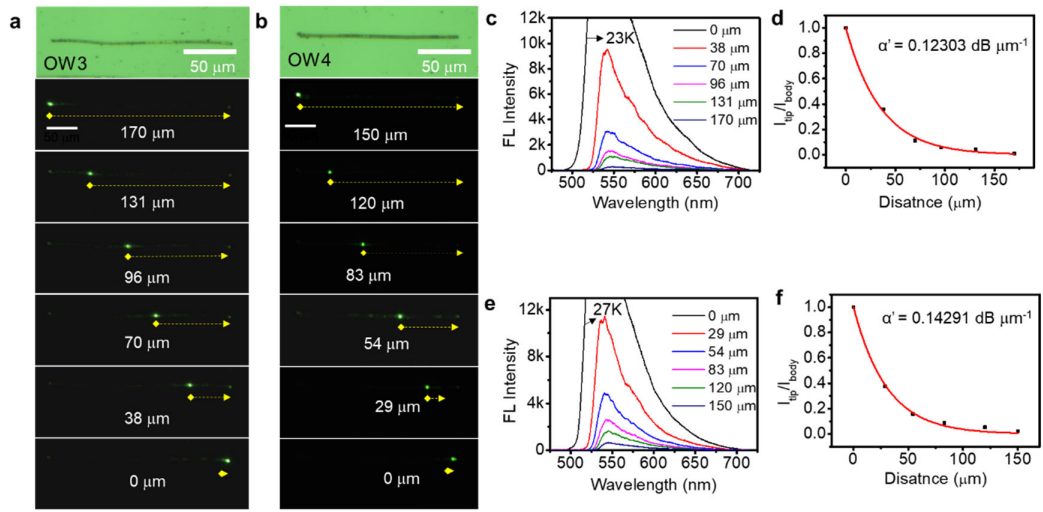

**Supplementary Figure 12. a, b** Confocal optical and FL microscope images of straight OW3 and OW4. **c, e** FL spectra collected at the right end of the waveguide for different optical path lengths in OW3 and OW4. **d, f** corresponding optical loss plots.

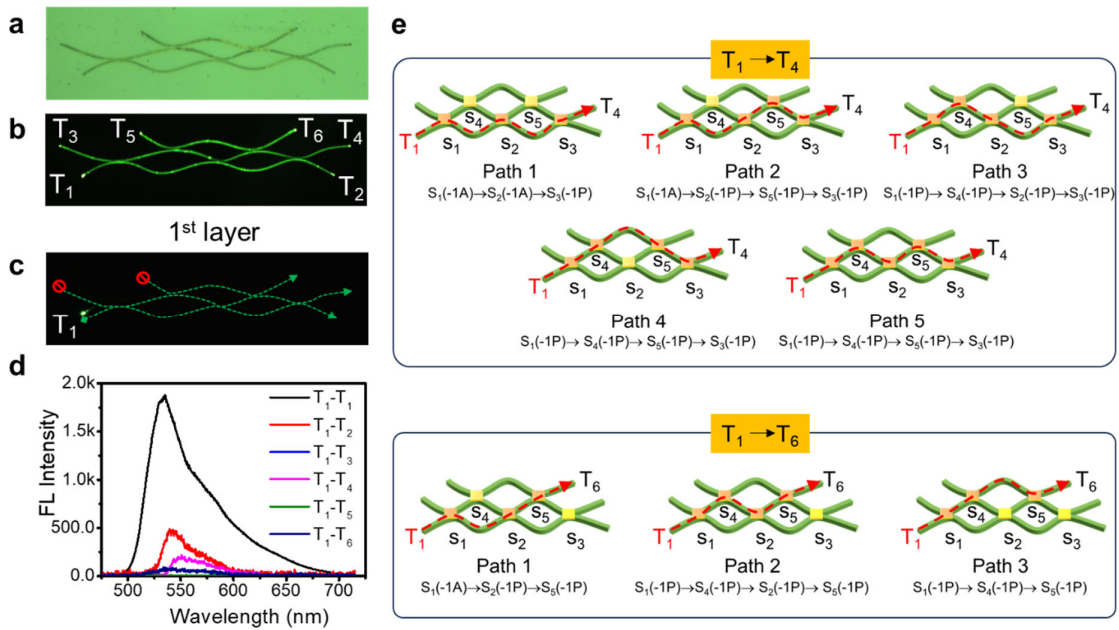

**Supplementary Figure 13. a, b** Confocal optical and FL microscope images of the three-layered ANN. **c** FL image of the three-layered ANN excited at $T_1$ terminal. **d** Corresponding FL spectra collected at all the terminals. **e** Graphical representation of three-layered ANN with five synapses showing different paths (red-dotted arrows) towards $T_4$, and $T_6$ for input at $T_1$.

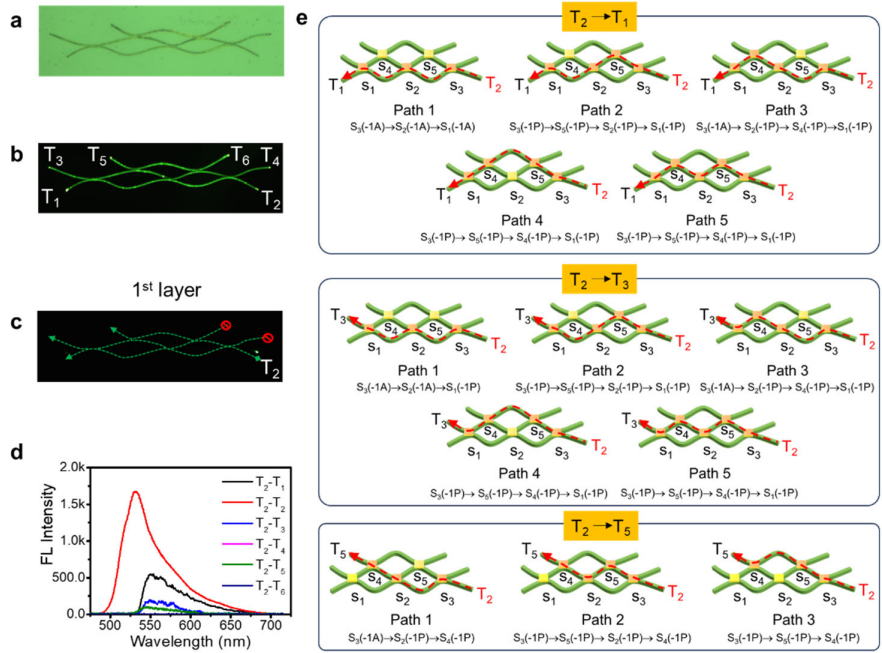

**Supplementary Figure 14. a, b** Confocal optical and FL microscope images of the three-layered ANN. **c** FL image of the three-layered ANN excited at $T_2$ terminal. **d** Corresponding FL spectra collected at all the terminals. **e** Graphical representation of three-layered ANN with five synapses showing different paths (red-dotted arrows) towards $T_1$, $T_3$, and $T_5$ for input at $T_2$.

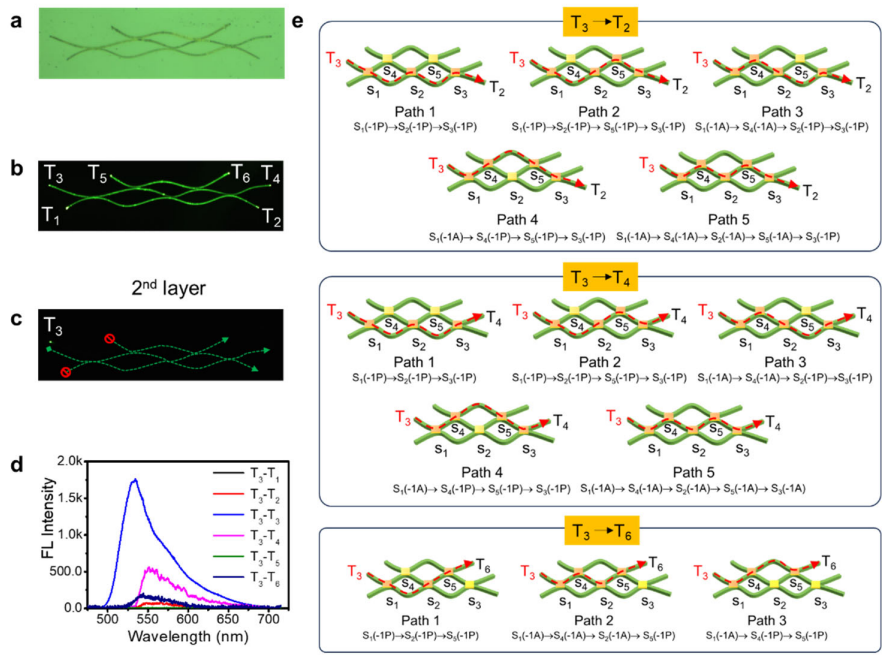

**Supplementary Figure 15. a, b** Confocal optical and FL microscope images of the three-layered ANN. **c** FL image of the three-layered ANN excited at $T_3$ terminal. **d** Corresponding FL spectra collected at all the terminals. **e** Graphical representation of three-layered ANN with five synapses showing different paths (red-dotted arrows) towards $T_2$, $T_4$, and $T_6$ for input at $T_3$.

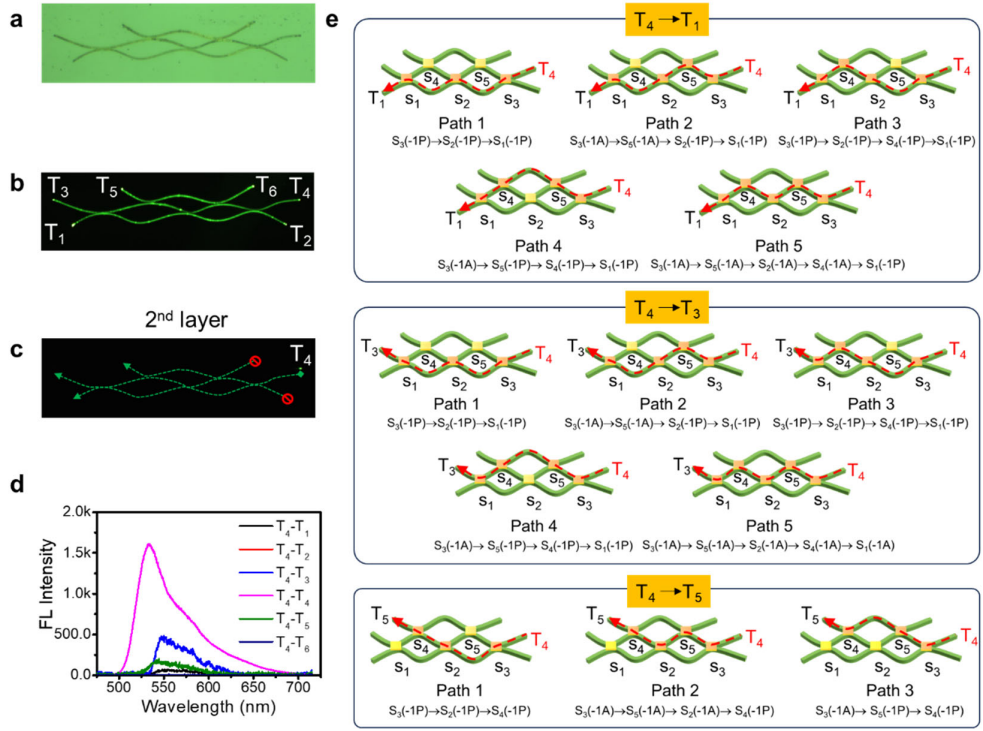

**Supplementary Figure 16. a, b** Confocal optical and FL microscope images of the three-layered ANN. **c** FL image of the three-layered ANN excited at $T_4$ terminal. **d** Corresponding FL spectra collected at all the terminals. **e** Graphical representation of three-layered ANN with five synapses showing different paths (red-dotted arrows) towards $T_1$, $T_3$, and $T_5$ for input at $T_4$.

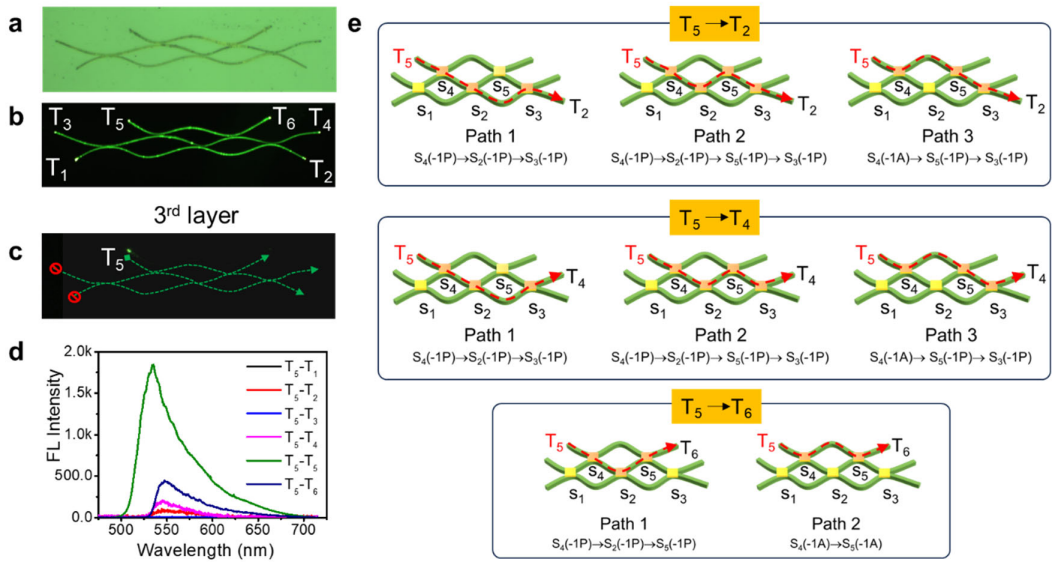

**Supplementary Figure 17. a, b** Confocal optical and FL microscope images of the three-layered ANN. **c** FL image of the three-layered ANN excited at $T_5$ terminal. **d** Corresponding FL spectra collected at all the terminals. **e** Graphical representation of three-layered ANN with five synapses showing different paths (red-dotted arrows) towards $T_2$, $T_4$, and $T_6$ for input at $T_5$.

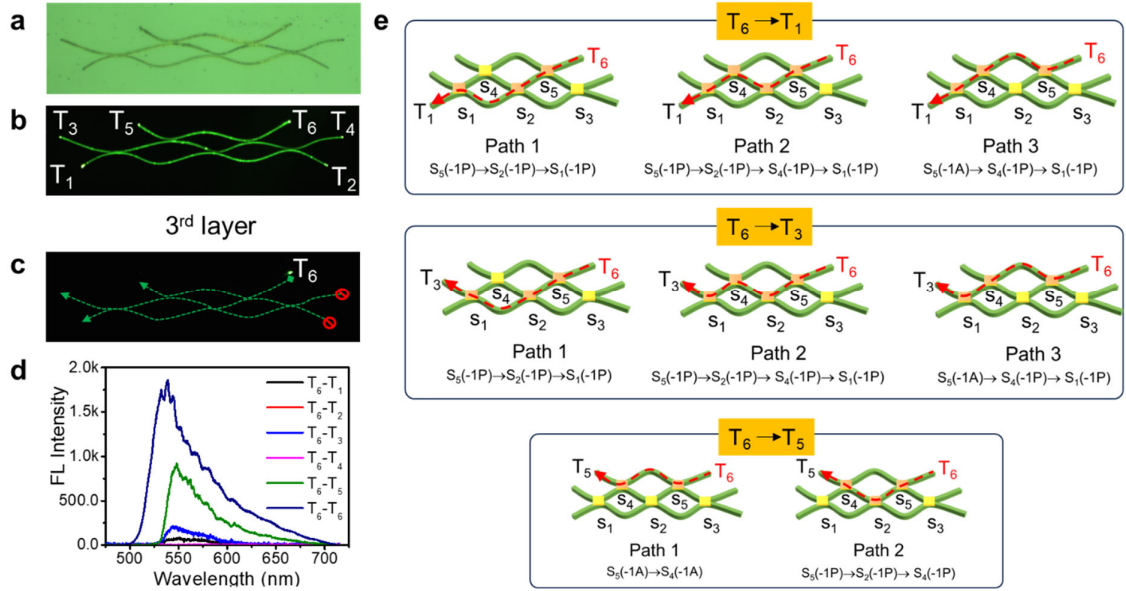

**Supplementary Figure 18. a, b** Confocal optical and FL microscope images of the three-layered ANN. **c** FL image of the three-layered ANN excited at $T_6$ terminal. **d** Corresponding FL spectra collected at all the terminals. **e** Graphical representation of three-layered ANN with five synapses showing different paths (red-dotted arrows) towards $T_1$, $T_3$, and $T_5$ for input at $T_6$.

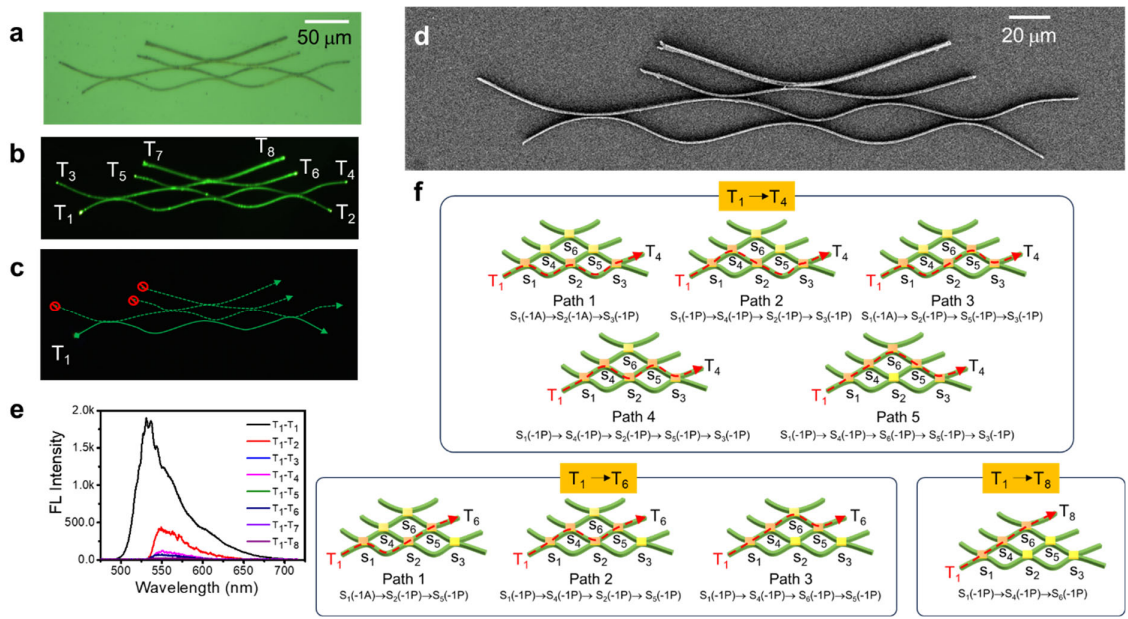

**Supplementary Figure 19. a, b** Confocal optical and FL microscope images of the four-layered ANN. **c** FL image of the four-layered ANN excited at T₁ terminal. **d** Corresponding FESEM image. **e** Corresponding FL spectra collected at all the terminals. **f** Graphical representation of four-layered ANN with six-synapses showing different paths (red-dotted arrows) towards T₄, T₆, and T₈ for input at T₁.

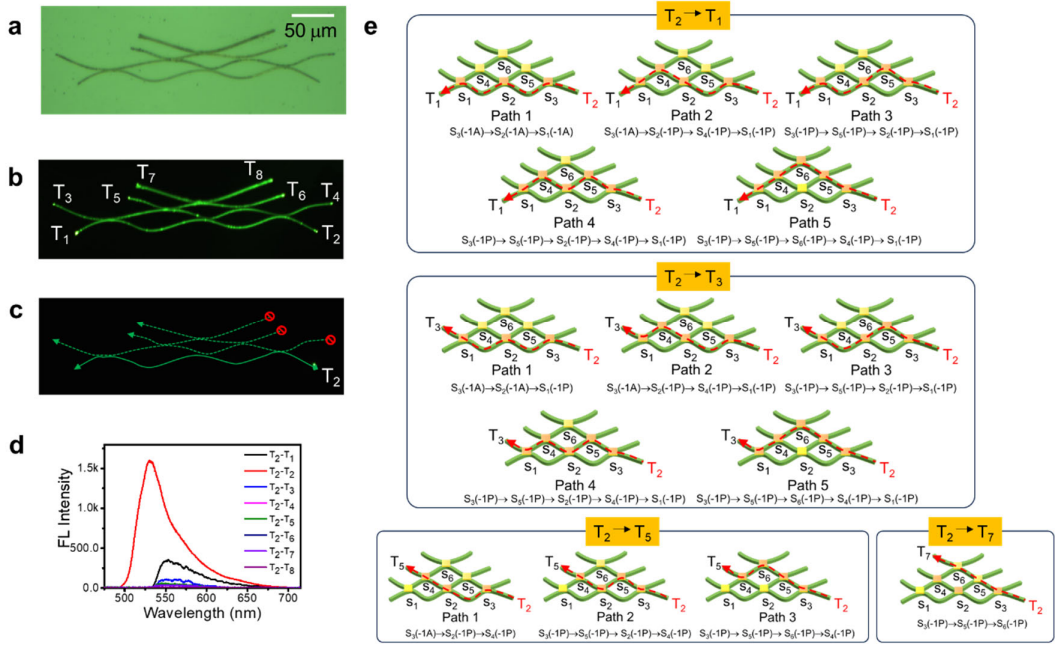

**Supplementary Figure 20. a, b** Confocal optical and FL microscope images of the four-layered ANN. **c** FL image of the four-layered ANN excited at T₁ terminal. **d** Corresponding FL spectra collected at all the terminals. **e** Graphical representation of four-layered ANN with six-synapses showing different paths (red-dotted arrows) towards T₁, T₃, T₅, and T₇ for input at T₂.

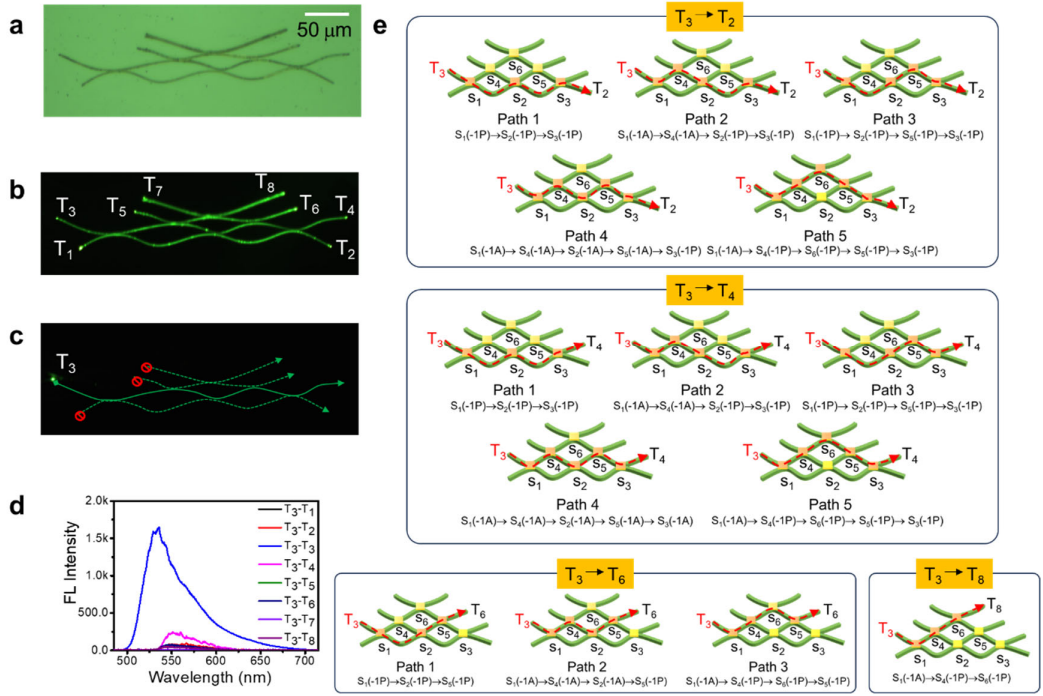

**Supplementary Figure 21. a, b** Confocal optical and FL microscope images of the four-layered ANN. **c** FL image of the four-layered ANN excited at $T_3$ terminal. **d** Corresponding FL spectra collected at all the terminals. **e** Graphical representation of four-layered ANN with six-synapses showing different paths (red-dotted arrows) towards $T_2$, $T_4$, $T_6$, and $T_8$ for input at $T_3$.

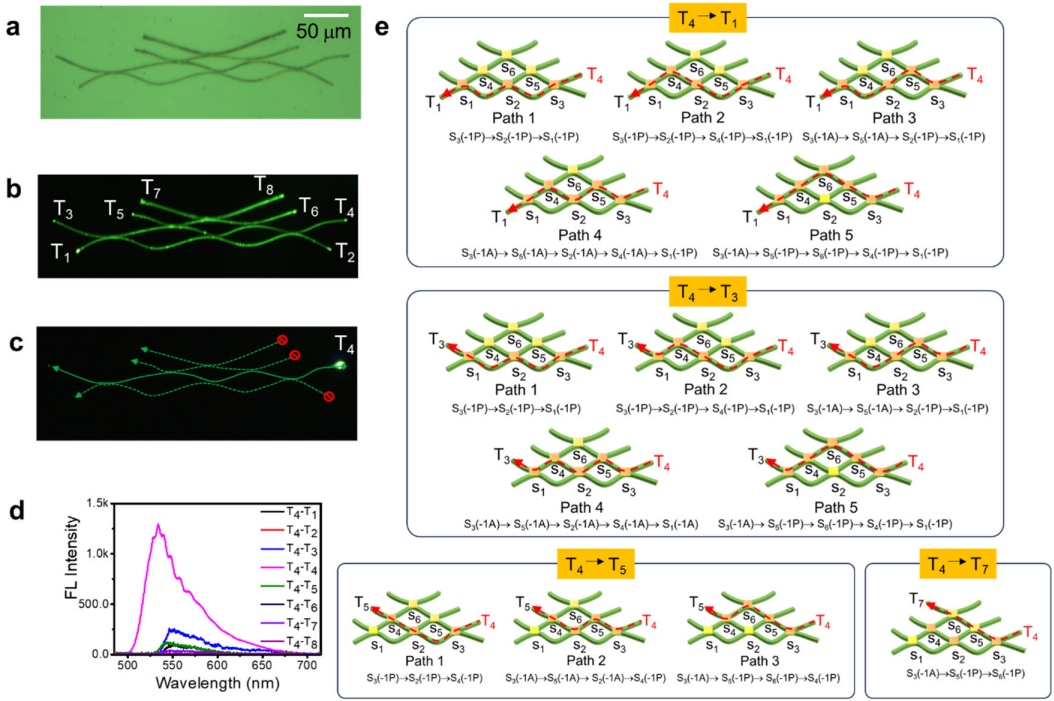

**Supplementary Figure 22. a, b** Confocal optical and FL microscope images of the four-layered ANN. **c** FL image of the four-layered ANN excited at $T_4$ terminal. **d** Corresponding FL

spectra collected at all the terminals. **e** Graphical representation of four-layered ANN with six-synapses showing different paths (red-dotted arrows) towards $T_1$, $T_3$, $T_5$, and $T_7$ for input at $T_4$.

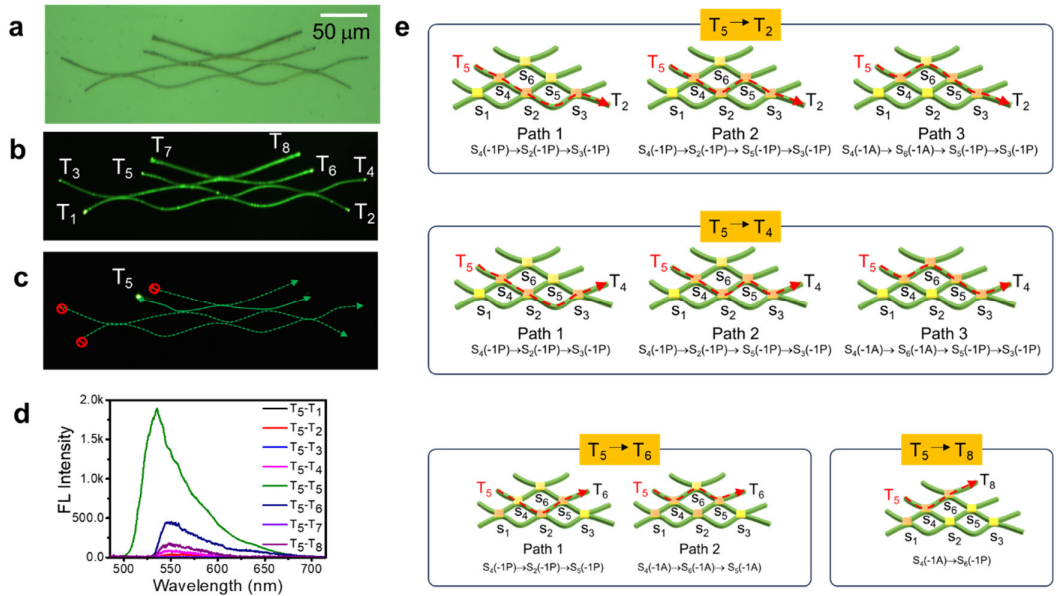

**Supplementary Figure 23. a, b** Confocal optical and FL microscope images of the four-layered ANN. **c** FL image of the four-layered ANN excited at $T_5$ terminal. **d** Corresponding FL spectra collected at all the terminals. **e** Graphical representation of four-layered ANN with six-synapses showing different paths (red-dotted arrows) towards $T_2$, $T_4$, $T_6$, and $T_8$ for input at $T_5$.

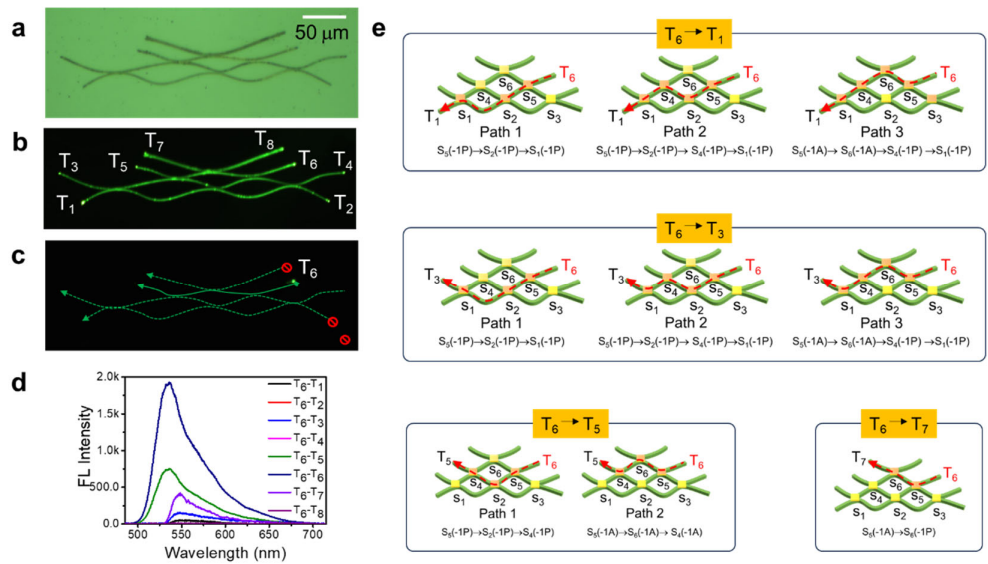

**Supplementary Figure 24. a, b** Confocal optical and FL microscope images of the four-layered ANN. **c** FL image of the four-layered ANN excited at $T_6$ terminal. **d** Corresponding FL spectra collected at all the terminals. **e** Graphical representation of four-layered ANN with six-synapses showing different paths (red-dotted arrows) towards $T_1$, $T_3$, $T_5$, and $T_7$ for input at $T_6$.

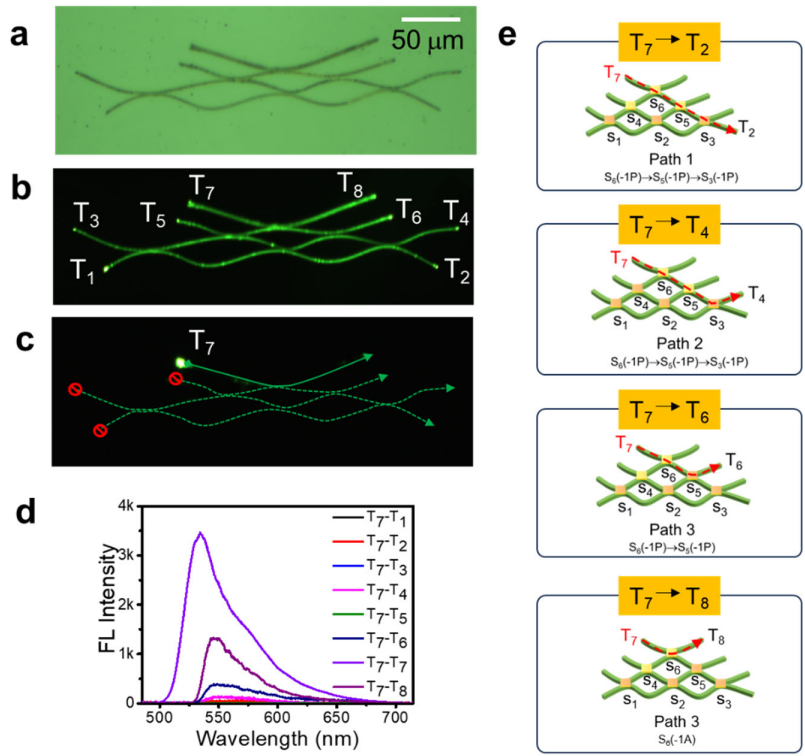

**Supplementary Figure 25. a, b** Confocal optical and FL microscope images of the four-layered ANN. **c** FL image of the four-layered ANN excited at $T_7$ terminal. **d** Corresponding FL spectra collected at all the terminals. **e** Graphical representation of four-layered ANN with six-synapses showing different paths (red-dotted arrows) towards $T_2$, $T_4$, $T_6$, and $T_8$ for input at $T_7$.

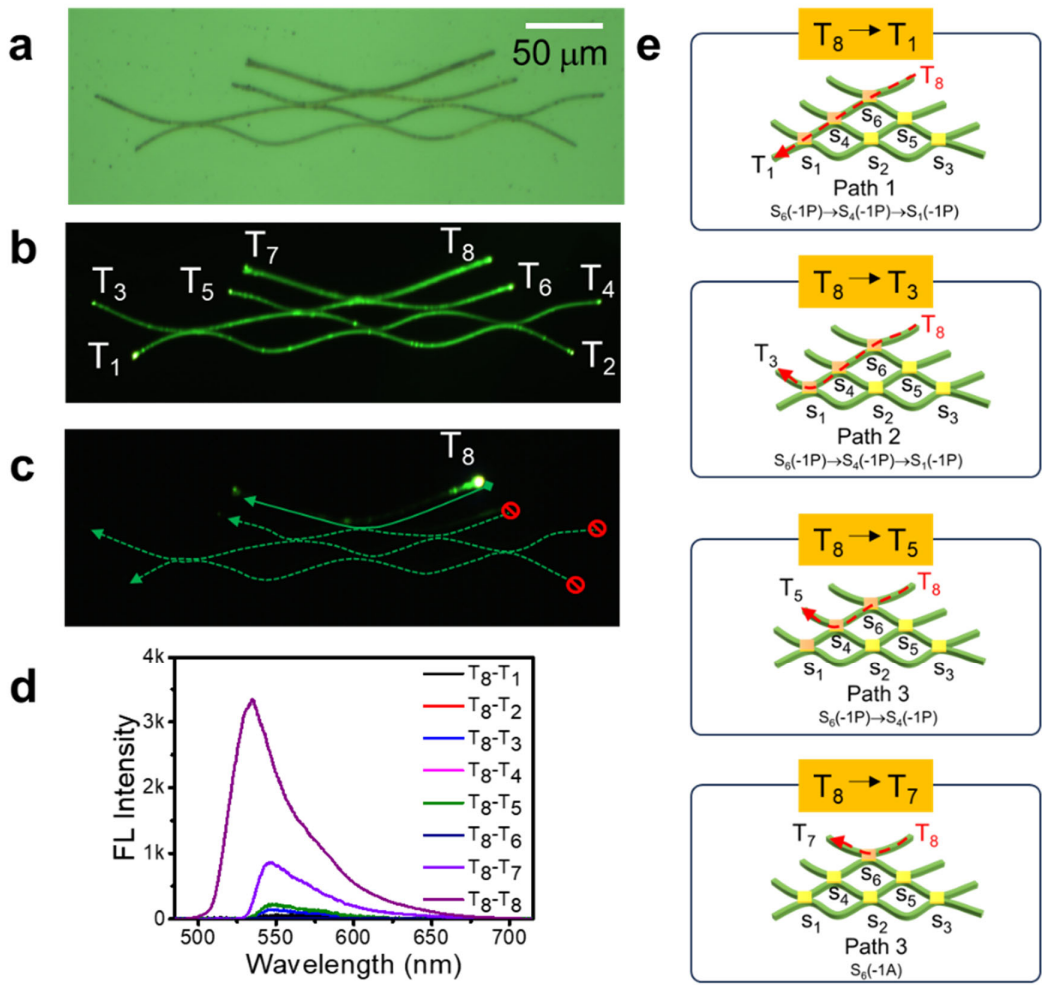

**Supplementary Figure 26. a, b** Confocal optical and FL microscope images of the four-layered ANN. **c** FL image of the four-layered ANN excited at $T_8$ terminal. **d** Corresponding FL spectra collected at all the terminals. **e** Graphical representation of four-layered ANN with six-synapses showing different paths (red-dotted arrows) towards $T_1$, $T_3$, $T_5$, and $T_7$ for input at $T_8$.

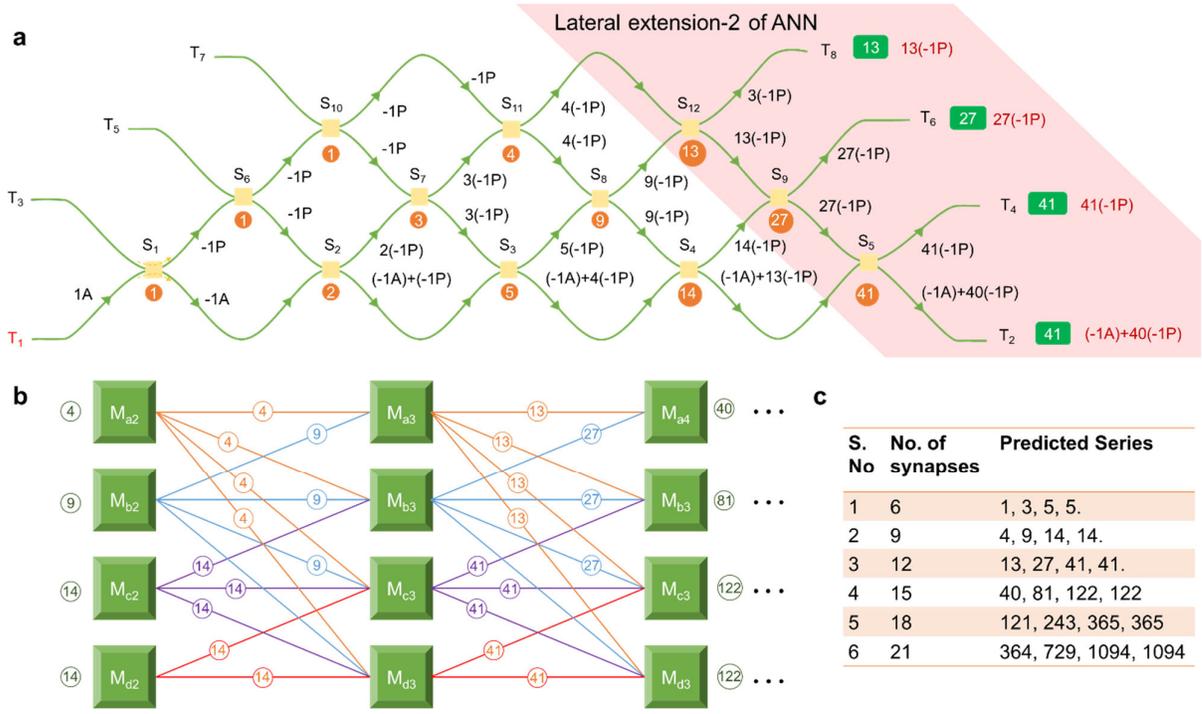

**Supplementary Figure 27. a** Mixing of signals in laterally extended four-level ANN with additional six synapses. **b** Prediction of number of mixed signals in the extended ANN from the previous set of data. M denotes mixed signals. Whereas, {an, bn, cn, dn} (n = integer) denotes the obtained mixed signals at the terminal $T_8$, $T_6$, $T_4$, and $T_2$, respectively. **d** Table depicting statistically predicted number of mixed signals that will be routed to outputs for incorporating more synapses.

**Supplementary note**
**Prediction of the next series of mixed signals in the extended ANN**
The computed set of values for the actual fabricated ANN were taken into account in order to estimate the mixed signals outcoupling at the output ports $T_2$, $T_4$, $T_6$, and $T_8$ of the laterally expanded ANN. Let $a_1$, $b_1$, $c_1$, and $d_1$, be the first set of values for the actual ANN, whereas $a_1$, $b_1$, $c_1$, and $d_1$, denotes the number of mixed signals obtained at $T_8$, $T_6$, $T_4$, and $T_2$, respectively. Lateral extension of the ANN will result in the formation of new synapses. By using the summing relationship of the preceding values, this results in the prediction of the subsequent set of mixed signals that emerge at the output ports.
For, example,
let's consider the first set of values ($a_1$, $b_1$, $c_1$, $d_1$) obtained from the actual ANN is (1, 3, 5, 5).

**Observations based on the sum rule:**
- 1 + 3 = 4 (Sum of the first two numbers in the first set gives the first number of the second set)
- 1 + 3 + 5 = 9 (Sum of the first three numbers in the first set gives the second number of the second set)

- 1 + 3 + 5 + 5 = 14 (Sum of all the numbers in the first set gives the third and fourth numbers of the second set)

So, the obtained mixed signals ($a_2$, $b_2$, $c_2$, $d_2$) for the extended ANN are (4, 9, 14, 14). (Supplementary figure. 27)

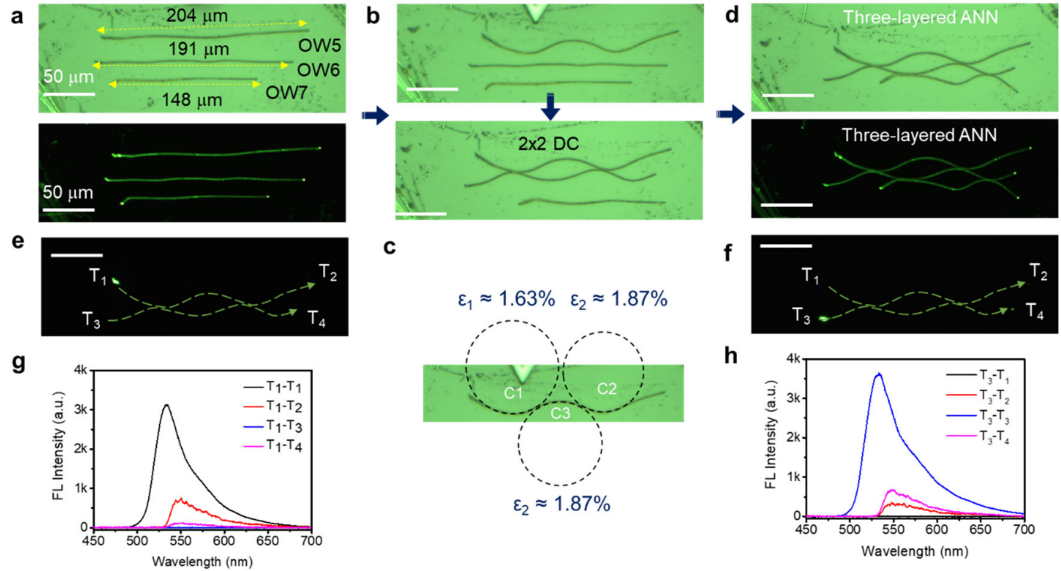

**Supplementary Figure 28. a** Confocal optical and FL images of a three straight single crystal optical waveguides OW5 – OW7 arranged parallel to each other, **b** mechanically bent OW5, bending of straight OW6 and creation of 2×2 directional coupler (DC), by integrating bent OW5 and OW6. **c** strain calculation at the curved regions in bent OW5 micromanipulated using AFM cantilever tip. **d** micromechanical bending and integration of OW7 with OW6 to form three-layered monolithic ANN. **e,f** FL images of DC excited at $T_1$ and $T_3$ terminals. The scale bar is 50 μm. **g,h** FL spectra recorded at other termini when excited at $T_1$ and $T_3$. The green dotted lines in FL images indicate the direction of light propagation in the 2×2 DC.

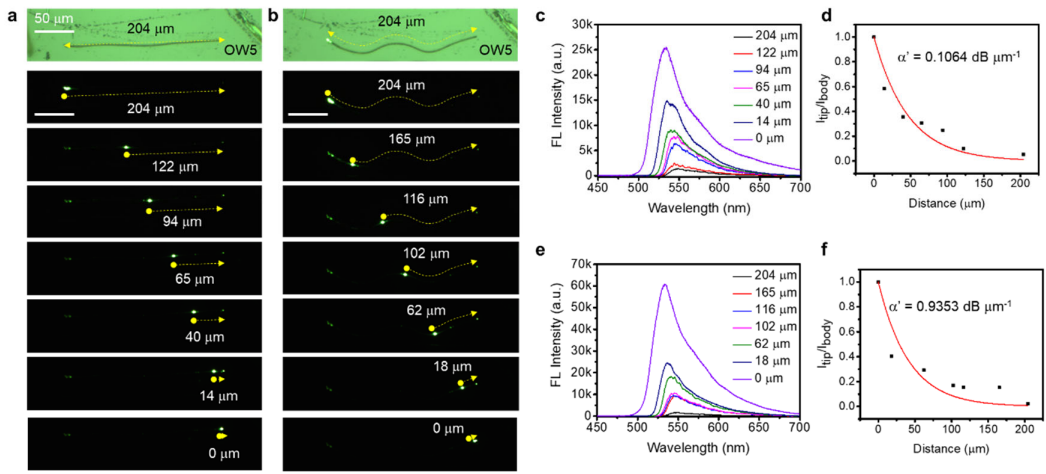

**Supplementary Figure 29.** Confocal optical and FL microscope images of **a** straight and **b** mechanically bent MPyIN crystal waveguide (OW5). FL spectra were collected at the right end

T$_2$ for multiple optical path lengths in **c** straight and **e** bent OW1. **d, f** corresponding optical loss plots. The α′ represents the optical loss coefficient for respective waveguides.

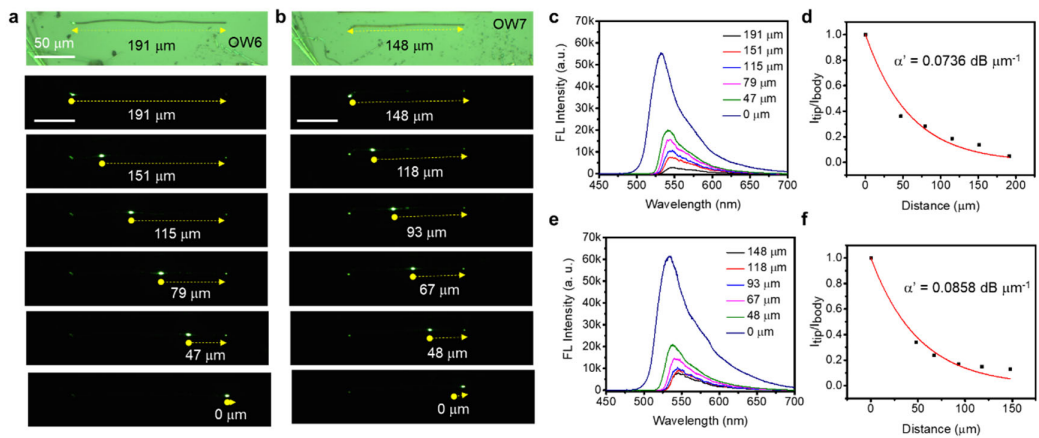

**Supplementary Figure 30. a, b** Confocal optical and FL microscope images of straight MPyIN crystal waveguides OW6 and OW7, respectively. FL spectra were collected at the right end for multiple optical path lengths in **c** OW6 and **e** OW7. **d, f** corresponding optical loss plots. The α′ represents the optical loss coefficient for respective waveguides.

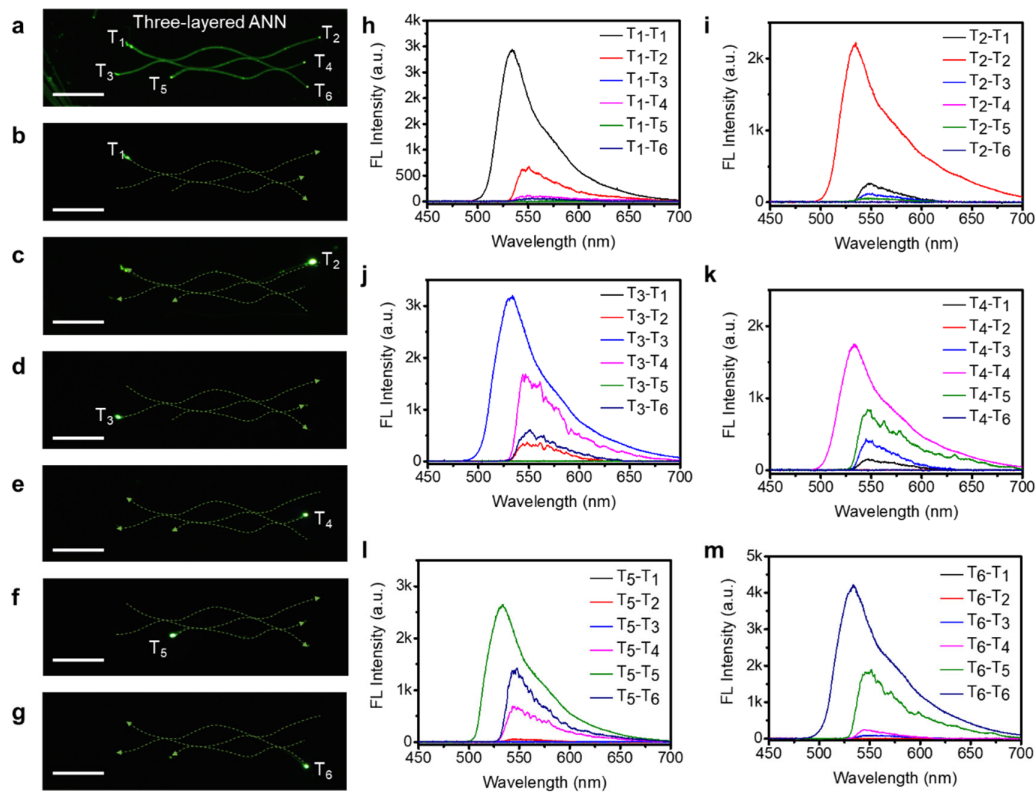

**Supplementary Figure 31. a** FL microscope image of the three-layered ANN. **b-g** FL images of the three-layered ANN excited at T$_1$-T$_6$ terminal. **h-m** Corresponding FL spectra collected at all the terminals.

## 5. Durability

The fabricated artificial neural network (ANN) remained stable for several months under ambient conditions. FESEM images of the ANN was captured two months post-fabrication.